\documentclass[pra,floatfix,amsmath,superscriptaddress,twocolumn]{revtex4}

\usepackage{amssymb}
\usepackage{graphicx}
\usepackage{graphics}
\usepackage{amsmath}
\usepackage{amsthm}
\usepackage{color}
\usepackage{bbding}
\usepackage{enumerate}

\def\bea{\begin{eqnarray}}
\def\eea{\end{eqnarray}}
\def\bi{\begin{itemize}}
\def\ei{\end{itemize}}
\def\bc{\begin{center}}
\def\ec{\end{center}}
\def\ba{\begin{aligned}}
\def\ea{\end{aligned}}
\def\be{\begin{equation}}
\def\ee{\end{equation}}
\def\bestar{\begin{equation*}}
\def\eestar{\end{equation*}}
\def\bt{\begin{tabular}}
\def\et{\end{tabular}}

\def\C{\hbox{$\mit I$\kern-.7em$\mit C$}} 
\def\R{\hbox{$\mit I$\kern-.6em$\mit R$}}

\def\Order{\mathcal O}
\def\one{\mbox{$1 \hspace{-1.0mm}  {\bf l}$}}

\def\ket#1{\left| #1\right>} 							
\def\bra#1{\left< #1\right|}							
\def\bk#1{\left< #1 \right>}							
\def\kb#1#2{\ket{#1}\bra{#2}}  							
\def\bmk#1#2#3{\left< #1\middle| #2 \middle| #3 \right>}  
\def\proj#1{\ket{#1}\bra{#1}}							

\DeclareMathOperator{\tr}{tr}

\DeclareMathOperator{\poly}{poly}

\def\Jmax{J_{\textrm{max}}}

\def\Bmax{B_{\textrm{max}}}
\def\sigmain{\sigma_{\mathrm{in}}}
\def\rhoin{\rho_{\mathrm{in}}}

\def\Psiin{\Psi_{\mathrm{in}}}
\def\xiin{\xi_{\mathrm{in}}}

\newcommand{\expect}[1]{\langle #1 \rangle}

\newcommand{\adj}[1]{#1^{\dagger}}

\newcommand{\Varepsilon}{\mathscr{E}}

\usepackage{color}
\usepackage[usenames,dvipsnames]{xcolor}
\usepackage{subfigure}
\newtheorem{theoremb}{Theorem}

\usepackage{float}
\usepackage{amssymb}
\usepackage[mathscr]{eucal}

\begin{document}

\author{W. L. Boyajian}
\affiliation{Institute for Theoretical Physics, University of
Innsbruck, Innsbruck, Austria}
\author{V. Murg}
\affiliation{Institute for Theoretical Physics, University of
Vienna, Vienna, Austria}
\author{B. Kraus}
\affiliation{Institute for Theoretical Physics, University of
Innsbruck, Innsbruck, Austria}
\title{Compressed Simulation of evolutions of the XY--model}

\begin{abstract}

We extend the notion of compressed quantum simulation to the $XY$--model. We derive a quantum circuit processing $\log(n)$ qubits which simulates the 1D XY--model describing $n$ qubits. In particular, we demonstrate how the adiabatic evolution can be realized on this exponentially smaller system and how the magnetization, which witnesses a quantum phase transition can be observed. Furthermore, we analyze several dynamical processes, like quantum quenching and finite time evolution and derive the corresponding compressed quantum circuit.

\end{abstract}
\maketitle

\section{Introduction}

Due to the exponential growths of the dimension of the Hilbert space as a function of the number of the constituting subsystems, the simulation of a quantum system with a classical computer seems to be unfeasible. Hence, classically the simulation of a quantum many--body system is often restricted to a very small number of constituents. However, the system can be simulated by another, better controllable, quantum system. The latter one is such that single--particle evolutions and the interactions between the subsystems can be realized sufficiently well. Moreover, the measurements can be performed very precisely. Such a quantum simulation was originally proposed by Feynman \cite{Fey82} in 1982. The fact that the evolution of one quantum system can indeed be simulated by another was then proven by LLoyd \cite{Llo96}. The suitability for the realization of such a quantum simulator has been shown, for instance, for experimental schemes based on ion--traps, or optical lattices \cite{PoCi04,JaZo05,PoCi08}. Due to the impressive experimental progress during the last decades, regarding the isolation of single particles, their manipulation and measurements, quantum simulations are performed in many experiments nowadays, among them experiments utilizing trapped ions, NMR, or atoms in optical lattices (see for instance \cite{Blo12,Lanyon11,Houck12} and references therein).

A quantum simulator can be employed to investigate several different physical phenomenons. It can be used to study the ground state properties of certain condensed matter systems and to observe quantum phase transitions \cite{QuantumPhaseTransitions}. In contrast to classical phase transitions, quantum phase transitions occur at zero temperature due to the change of some parameter, like the strength of the magnetic field, or pressure. Moreover, a quantum simulator enables us to determine the dynamical behavior of a quantum many--body system and processes like quantum quenching \cite{Zurek-KZMmechanism} can be investigated.
Preferable for certain experimental demonstrations of quantum simulations are models where the interactions between the constituting subsystems are quasi--local, i.e. involve only a few neighboring particles. Quantum spin models are very well suited for that. Examples of such models are the quantum Ising--model, or (more generally) the quantum XY--model, which we consider here. Even though in one dimension those models can be simulated classically efficiently, they are good candidates for the realization of quantum simulations, since they show interesting physical phenomenons. For instance, they display quantum phase transitions if the strength of the external magnetic field is varied. In addition to that, they provide insight in real physical systems.
For example, the XY model with field
is a good model for helium absorbed on metallic surface \cite{domb88}.
Furthermore, it gives the master equation of the kinetic Ising model \cite{siggia77}.

One approach to observe the quantum phase transition of for instance the quantum Ising model is the following. The system, consisting of $n$ qubits, is initially prepared in the ground state of the Hamiltonian describing the interaction between the subsystems without an external magnetic field. Then, a quantum algorithm, which is based on the adiabatic theorem, transforms this input state to the ground state of the Hamiltonian corresponding to a non--vanishing external magnetic field. Measuring then the magnetization of the system as a function of the strength of the magnetic field displays the phase transition in case $n$ is sufficiently large. Note that this simulation can either by performed in a digital or an analog fashion. We will focus here on the digital simulation, where specific unitary operations are applied in order to force the system to evolve into the ground state of the Hamiltonian with an increased magnetic field. In this way the ground state properties of the Ising model for $n=2$ \cite{FrScNatPhys08} and for $n=3$ \cite{KiCh10} has been recently observed experimentally.

In \cite{Kraus-matchgates} one of us presented a different approach to observe the quantum phase transition. There, the adiabatic evolution was not directly simulated, but an exponentially smaller quantum system was employed to reproduce it. More precisely, it is shown there, how the phase transition of the $1$D Ising model of a spin chain consisting out of $n$ qubits can be observed via a compressed algorithm running only on $\log(n)$ qubits. The main reason for this to be possible is that the adiabatic evolution described above including the measurement of the magnetization can be described by a matchgate circuit. In \cite{JozsaMiyake-matchgates} it has been shown that such a circuit can be simulated by a universal quantum computer processing only $\log(n)+3$ qubits. The number of basic operations, i.e. the number of single and two--qubit gates which are required to implement the algorithms, coincides for both algorithms. For the simulation of the Ising model it could be shown that the symmetry of the model allows to compress the simulation even to $\log(n)$ qubits. To give an example, consider an experimental set--up, where up to $8$ qubits can be well controlled. Such a set--up could be used to simulate the interaction of $2^8=256$ qubits. For such a large system the phase transition can be well observed.

The aim of this paper is on the one hand to generalize this result to the XY--model, and on the other hand to show that various other interesting processes, like quantum quenching, where the evolution is non--adiabatic, can be simulated with an exponentially smaller system. Furthermore, we also demonstrate how the time evolution governed by the XY--model can be simulated. Moreover, we will show that the number of elementary gates, i.e. single and two--qubit gates required for the simulation can be even smaller in the compressed simulation than the original one.

The remainder of the paper is organized as follows. In Section \ref{sec:preliminaries} we introduce our notation and review the notion of matchgate circuits. Then, we recall two theorems, dealing with the classical as well as the quantum simulation of a matchgate circuit. Moreover, we recall the concepts of adiabatic evolution and quantum quenching and review some basic properties of the XY--model. In Section \ref{sec:magnetization} we present a matchgate circuit, which simulates the adiabatic evolution and derive the corresponding compressed quantum circuit, which involves only $\log(n)+1$ qubits. Then, we show how the symmetry of the XY--interaction can be used to compress the whole simulation into $\log(n)$ qubits. The number of single and two--qubit gates required in the compressed circuit will be shown to be much smaller than in the original one. In the subsequent section (Sec. \ref{sec:quenching}) we show how quantum quenching can be simulated with an exponentially smaller system. Finally, in Sec. \ref{sec:timeevolution} we derive a compressed simulation of the time evolution of the system.

\section{Preliminaries}\label{sec:preliminaries}

In this section, we introduce our notation and recall some results on matchgate circuits which are relevant for the present work. After that, we summarize some basic facts regarding the XY--model.

\subsection{Notation}

The computational basis is denoted by $\ket{k}$ for $k\in \{0,1\}^{n}$. We use the standard notation to denote the Pauli operators, i.e.
$X=\ket{0}\bra{1}+\ket{1}\bra{0}$, $Y=-i\ket{0}\bra{1}+i\ket{1}\bra{0}$, $Z=\ket{0}\bra{0}-\ket{1}\bra{1}$. The states $\ket{+}$, $\ket{-}$, $\ket{+_y}$, $\ket{-_y}$ will represent the eigenstates with eigenvalues $1$ and $-1$ of the operators $X$ and $Y$ respectively. The tensor product symbol will be
omitted whenever it does not cause any confusion and $\one$ will
denote the identity operator. For any quantum circuit, its
{\em size}, $N$, is its total number of single and two--qubit gates, and its {\em width}, $n$, is the total number of qubits upon which its gates act.
Throughout the paper $n$ is assumed to be a power of 2 and the logarithm is taken in base 2.

\subsection{Matchgate circuits and their classical and quantum simulation} \label{sub: matchgates and theorems}

Matchgates constitute an interesting class of two--qubit gates, which occur for example in
the theory of perfect matchings of graphs, non-interacting
fermions, and one-dimensional spin chains. A \emph{matchgate} is a 2-qubit gate $G(A,B)$ of the form

\bea
G(A,B)=
\left(
\begin{array}{rrrr}
    p & 0 & 0 & q \\
    0 & w & x & 0 \\
    0 & y & z & 0 \\
    r & 0 & 0 & s \\
\end{array}
\right),
\label{eq:Matchgate}
\eea
where

\begin{center}
\begin{tabular}{c c}
	$A=
	\left(
	\begin{array}{rr}
	    p & q \\
	    r & s \\
	\end{array}
	\right)$
&
	and	$B=
	\left(
	\begin{array}{rr}
	    w & x \\
	    y & z \\
	\end{array}
	\right)$,

\end{tabular}
\end{center}
are unitaries which have the same determinant.

We will call a \emph{matchgate circuit} a quantum circuit consisting solely of gates like the one in Eq. (\ref{eq:Matchgate}). Two important results, which are relevant here, have been proven for matchgate circuits where each gate is acting on nearest neighbors (n.n), the input states are computational basis states and the outcome is obtained via a single qubit $Z$--measurement \cite{JozsaMiyake-matchgates,JozsaKraus-matchgates}. The first result shows that any of those circuits can be computed efficiently on a classical computer (see Theorem \ref{thm:efficientsimulation} below). The second result establishes an equivalence between matchgate circuits and universal quantum computation running on exponentially less qubits (see Theorem \ref{thm:compression} below). We recall now those theorems, which have been proven in \cite{JozsaMiyake-matchgates} and \cite{JozsaKraus-matchgates}.

\begin{theoremb}\label{thm:efficientsimulation} \cite{JozsaMiyake-matchgates}: Consider a matchgate circuit of size $N$ and width $n$ such that

\begin{enumerate}[(i)]
	\item   the input state is any computational basis state, $\ket{x_1...x_n }$,
    \item   it comprises $N$ nearest neighbors matchgates,
    \item   and the output is a final measurement in the computational basis of some single qubit $k$.
\end{enumerate}
Then, the output can be computed classically in $\textrm{poly}(n,N)$ steps. \end{theoremb}
That is, for any qubit $k$ the expectation value $\langle Z_k \rangle_{\textrm{out}}$ can be computed classically efficiently. We recall the proof of this theorem below. Note that it is important to restrict the circuit to n.n. matchgates. In fact, in \cite{JozsaMiyake-matchgates} it has been shown that allowing in addition to the n.n. matchgates next nearest neighbor matchgates, would be sufficient to obtain universal quantum computation. Since we consider here only n.n. matchgates, we omit the n.n. specification in the following when referring to gates or circuits.

In order to state the second result we denote by $MG(n,N)$ a matchgate circuit acting on $n$ qubits which obeys the conditions (i--iii) above and has size $N$, i.e. it is composed of $N$ matchgates. $QC(m;M)$ denotes a universal quantum circuit acting on $m$ qubits of size $M$, whose output is also a measurement in the computational basis of a single qubit. We call two circuits equivalent if they simulate each other and compute the same outcome. Using this notation, we can now state the following theorem.

\begin{theoremb}\label{thm:compression} \cite{JozsaKraus-matchgates}: The following equivalence holds.
\begin{enumerate}[(a)]
	\item	Given a matchgate circuit $MG(n;N)$ there exists an equivalent quantum circuit $QC(m;M)$ with $m=\lceil \log{(n)} \rceil +3$ and $M=\Order[N\log{(n)}]$. The encoding of the circuit $QC$ can be computed from the encoding of the circuit
$MG$ by means of a (classical) space ${\cal O}(\log n)$ computation.
    \item	Given a quantum circuit $QC(m;M)$ there exists an equivalent matchgate circuit $MG(n;N)$ with $n=2^{m+1}$ and $N=\Order(M2^{2m})$. The encoding of the circuit $MG$ can be computed from the encoding of the circuit
$QC$ by means of a (classical) space ${\cal O}(m)$ computation.
\end{enumerate}\end{theoremb}
Note that in both cases, the number of classical operations that are required to encode the gates from one circuit into the other is bounded by $\mathcal{O}(\log{n})$ and $\mathcal{O}(m)$ respectively. This fact excludes the possibility that the computation is performed by the classical computer. Hence, the equivalent quantum circuits occurring in this Theorem are indeed simulations of each other.
Thus, the computational power of, for instance, a polynomial--sized matchgate circuits and universal polynomial-sized quantum circuits running on $\log(n)$ qubits is equivalent (up to a classical log-space computation).

Let us review here the outline of the proofs of the previous theorems, since we will need some of these details later on. Considering $n$ qubits, we introduce the set of $2n$ hermitian operators, $\{c_j\}$, which satisfy the anticommutation relations (see for instance \cite{JozsaMiyake-matchgates})
\be
	\{c_j,c_k\}=c_jc_k+c_kc_j=2\delta_{j,k}\one,\quad j,k=1,\dots,2n .
	\label{eq:commutation relations}
\ee
That is, the operators $c_k$ define a \emph{Clifford Algebra} \footnote{Any element of this Algebra is obtained as a complex linear combinations of products of the generators $c_j$, i.e. an arbitrary element of the Algebra is of the form $\sum_{i_1<\dots <i_k }{A_{i_1\dots i_k}c_{i_1}\dots c_{i_k}}$.}. We are going to use the \emph{Jordan Wigner representation} to write the generators of the Algebra in terms of the Pauli matrices. In this representation, the generators $c_j$ are given by
\be
	\ba
		c_1=X I\dotsm I \quad  	& \dotso  & c_{2k-1}=Z\dotsm Z X I \cdots I \quad 	& \dotso  \\
		c_2=Y I\dotsm I\quad 	& \dotso  &  c_{2k}=Z\dotsm Z Y I \cdots I \quad 	& \dotso\\
	\ea,
	\label{eq:Clifford algebra}
\ee
where the operators $X$ and $Y$ act on the $k$-th slot in $c_{2k-1}$ and $c_{2k}$, and $k$ ranges from 1 to $n$. A Hamiltonian is said to be \emph{quadratic} in the elements of the Clifford algebra if it can be written as
\be
	H=i \sum_{j \neq k=1}^{2n}{h_{j,k}c_jc_k},
	\label{eq:Quadratic Hamiltonian}
\ee
where $h$ is a real antisymmetric matrix. Note that the operators $c_k$  change in a particular way under conjugation by unitaries of the form of $U=e^{-i\alpha H}$, where $H$ is given as in Eq. (\ref{eq:Quadratic Hamiltonian}). In \cite{JozsaMiyake-matchgates} it was proven that
\be
	U^{\dagger}c_jU=\sum_{k=1}^{2n}{R_{j,k}c_k},
	\label{eq:UcU}
\ee
where $R=e^{4\alpha h}\in \mathcal{SO}(2n)$ is a real $2n \times 2n$ matrix.

The connection with matchgates is obtained by the fact that any matchgate, i.e. any gate of the form of Eq. (\ref{eq:Matchgate}), or any product of matchgates, corresponds to the evolution of a quadratic Hamiltonian. The inverse also holds, i.e. any unitary which can be written as the exponential of a quadratic Hamiltonian can be decomposed into matchgates \cite{JozsaKraus-matchgates}.

Using Eq. (\ref{eq:UcU}), Theorem 1 can be easily proven as follows. Consider a matchgate circuit, $MG(n,N)$, which is described by a unitary $U=U_N\cdots U_1$, where $U_i$ represents an arbitrary matchgate. Each unitary $U_i$ is associated to an orthogonal matrix $R_i$ according to Eq. (\ref{eq:UcU}). Then, the matrix $R$ associated to the whole circuit is given by $R=R_N\cdots R_1$, as can be easily seen from Eq. (\ref{eq:UcU}).
Let $\ket{\Psiin}$ denote the input state of the matchgate circuit. Using Eq. (\ref{eq:UcU}) and the fact that $Z_k=-ic_{2k-1}c_{2k}$, the expectation value of the $Z_k$ is given by
\be
	\ba
		\bk{Z_k}&=\bmk{\Psiin}{U^{\dagger}(-ic_{2k-1}c_{2k})U}{\Psiin}\\
				&=\sum_{j,l}{R_{2k-1,j}R_{2k,l}\bmk{\Psiin}{-ic_{j}c_{l}}{\Psiin}}\\
				&=[RSR^T]_{2k-1,2k}.
	\ea
\label{eq:Expectation Zk2}
\ee
Here, $S$ is defined by $S_{j,l}= \bmk{\Psiin}{-ic_j c_l }{\Psiin}$ for $j\neq l$ and $S_{j,j}=0$. Note that both matrices $R$ and $S$ are of dimension $2n\times 2n$. For instance, for the input state $\ket{\Psiin}=\ket{0}^{\otimes{n}}$, $S$ is given by
\be
	S=\one \otimes iY,
	\label{eq:S}
\ee
where $\one$ denotes the identity matrix of $n$ dimensions. Since the matrix $R$ is the product of $N$ matrices $R_i$ associated to each matchgate of the circuit, the number of operations required to compute it, scales polynomially in $N$ and $n$. Regarding the computation of $S$, note that the product $c_jc_l$ can be written as a product of Pauli operators [see Eq. (\ref{eq:Clifford algebra})]. Hence, in the particular case where the input state is a product state in the computation basis, the number of operation required to compute the matrix $S$ is a polynomial of $n$. Therefore, it is possible to compute the expectation value of $Z_k$, Eq. (\ref{eq:Expectation Zk2}), in $\mathrm{poly}(N,n)$ steps. Thus, the matchgate circuit can be simulated classically efficiently \footnote{The previous result holds for an arbitrary computational basis state as input state. However, it has been shown \cite{JozsaKraus-matchgates} that one can always choose without loss of generality the input state $\ket{\Psiin}=\ket{0}^{\otimes{n}}$.}.

The idea behind the proof of the first proof of Theorem \ref{thm:compression} is that the expectation value of for instance $\bk{Z_1}$ [see Eq. (\ref{eq:Expectation Zk2})] can be obtained by a circuit which acts only on $\log(n)+1$ qubits \footnote{Note that one can always choose $k=1$ without loss of generality \cite{JozsaKraus-matchgates}.}. There, the controlled unitary $\Lambda U=\kb{0}{0}\otimes \one + \kb{1}{1}\otimes U$ with $U=S^{-1}RSR^T$ unitary, is applied to the input state $\ket{+}\ket{0}^{\otimes\log{n}}$. Denoting by $\ket{k}$ the $k$-th element of the computational basis of a $2n$--dimensional Hilbert space [$\log(n)+1$ qubits], one can easily show that the expectation value of $X_1$ is $\bmk{1}{RSR^T}{2}$, which coincides with $\langle Z_1 \rangle$ in Eq. (\ref{eq:Expectation Zk2}). It is important to note that the conversion of the matchgates into their respective gates in the compressed circuit is performed by a classical computer which is bounded to ${\cal O}\big[\log(n)\big]$--space, which ensures that the computation is indeed performed by the quantum computer and not the classical one.

\subsection{Ising and XY model}

The 1D \emph{Ising} model or the more generally, the 1D \emph{XY model} describes a one-dimensional chain of spins with simple nearest neighbor interactions, and with an external magnetic field. As mentioned before, those models exhibit quantum phase transitions. Among others, these two ingredients make them interesting models for their experimental realization.

In this subsection we review some known results of the 1D XY--model, for open, periodic, and Jordan-Wigner (JW) boundary conditions \cite{lieb61, QuantumPhaseTransitions,katsura62}. Although in the limit of infinitely many spins, all the boundary conditions are equivalent, we briefly discuss finite size effects (see also Appendix A).

The $XY$--Hamiltonian governing the evolution of a 1D spin chain of $n$ qubits with \emph{open} boundary conditions is given by
\be
	H=-B\sum_{i=1}^{n}{Z_{i}}-J\sum_{i=1}^{n-1}{\left(X_{i}X_{i+1}+\delta Y_{i}Y_{i+1}\right)}.
	\label{eq:Introducing the Hamiltonian}
\ee
The first term describes the global magnetic field in $z$ direction, while the second and third term describe the interactions between n.n. spins in the $x$ and $y$ directions respectively and $\delta \in [0,1]$ denotes the anisotropy. In the Ising model only $x-x$ interactions occur, i.e. $\delta=0$. By choosing $\delta=1$ one obtains the Hamiltonian of the so--called \emph{XX model}. Those two cases are qualitatively different to those where $\delta\neq 0,1$ (see also Appendix A), but in the circuits presented in this paper, $\delta$ can be set to any value in the interval $[0,1]$.

In the following we write the Hamiltonian in Eq. (\ref{eq:Introducing the Hamiltonian}) as
\be
	H=-B H_0-J \left(H_1 + \delta H_2 \right),
	\label{eq:Hamiltonian simplified}
\ee
with
\be
	H_0=\sum_{i=1}^{n}{Z_{i}},\; H_1=\sum_{i=1}^{n-1}{X_{i}X_{i+1}},\; H_2=\sum_{i=1}^{n-1}{Y_{i}Y_{i+1}}.
	\label{eq:H0,H1,H2}
\ee

To obtain the Hamiltonian describing \emph{periodic} boundary conditions the interaction between the first and the $n$-th spins has to be added, i.e.
\be
	\bar{H}=H-J\left(X_{n}X_{1}+\delta Y_{n}Y_{1}\right),
	\label{eq:ClosedBC}
\ee
with $H$ given in Eq. (\ref{eq:Introducing the Hamiltonian}). Similarly, the Hamiltonian describing JW boundary conditions can be written as
\be
	\widehat{H}=H-J
	\left(X_{n} \tilde{Z} X_{1}+ \delta Y_{n} \tilde{Z} Y_{1} \right),
	\label{eq:JWBC}
\ee
where $\tilde{Z}=  \otimes_{i=1}^{n}{Z_i}$. Defining $X_{n+1}=\tilde{Z}X_1$ and $Y_{n+1}=\tilde{Z}Y_1$, this Hamiltonian can also be written as
\be
	 \widehat{H}\equiv -B\widehat{H}_{0}-J\left(\widehat{H}_{1} + \delta \widehat{H}_{2} \right),
	 \label{eq:HamiltonianJW simplified}
\ee
where, $\widehat{H}_{j}$ are defined analogously as $H_j$ in Eq. (\ref{eq:H0,H1,H2}), but where the sum runs form $1$ to $n$.

In the subsequent sections, we will focus on open and JW boundary conditions. The reason for that is that the corresponding evolutions correspond to matchgate circuits, as we will show below. Note that the choice between open and JW boundary conditions
is not going to affect the way the matchgate circuits are constructed. Thus, we will omit the hat in the Hamiltonian in order to refer indistinctly to both, open and JW boundary conditions, unless stated otherwise.

\subsection{Quantum phase transition}

To introduce the concept of quantum phase transition in the context of this paper, we consider a Hamiltonian of the form $H(s)$ where $s$ is a dimensionless parameter. If we consider a finite number of systems in the spin model, it could happen that the ground state energy changes as a smooth function of $s$, or that at some $s=s_c$ some excited state becomes the ground state, creating a point of non analyticity in the ground state energy \cite{QuantumPhaseTransitions}. A point like this is called a \emph{level-crossing}. When an infinite number of systems are considered, an additional situation may appear, since energy levels which did not cross in the finite case (\emph{avoided} level-crossing) may get closer as the number of systems increases and finally overlap in the asymptotic limit. In this case, the ground state energy also becomes non-analytic at this point, and in both cases one says that the system exhibits a quantum phase transitions \cite{QuantumPhaseTransitions}. Unlike the classical phase transitions, the quantum counterpart occurs at zero temperature, and its nature is purely quantum.
For the Hamiltonians considered in this paper, the behavior of certain observables, like the magnetization is very abrupt (or discontinuous) when the system crosses through a critical point. Due to this characteristic, we can observe the phase transition by measuring the magnetization as a function of $s$. In Sec \ref{sec:magnetization} we will derive a matchgate circuit which can be used for this purpose.

\subsection{Adiabatic evolution and quantum quenching}\label{sec:Adiabatic}

\emph{Adiabatic evolution} and \emph{quantum quenching} are two processes where the evolution of a quantum system is governed by a Hamiltonian that changes continuously in time. The difference between the two evolutions lies in the speed with which the Hamiltonian is changed. For adiabatic changes, a system which is initially prepared in a ground state of the Hamiltonian will (under certain conditions) remain in the ground state (adiabatic theorem). However, if the change is too abrupt, excitations will occur. We recall the basic ideas of adiabatic evolution and quantum quenching here since we are going to present later on (see Secs. \ref{sec:magnetization},\ref{sec:quenching} and \ref{sec:timeevolution})) for both scenarios a compressed quantum algorithm which can simulate the original evolution.

The \emph{Adiabatic Theorem} relates the speed of the time evolution necessary for the system staying in the ground state to
the energy gap between the ground state and the first excited
state of the same symmetry~\cite{born28,kato50,friedrichs55}.
Assuming that the time evolution is determined by
the Hamiltonian $H(s) = \sum_j E_j(s) \proj{\Psi_j(s)}$
with eigenvalues $E_j(s)$ and eigenstates $\ket{\Psi_j(s)}$
which continuously turns from $H(0)$ at the time $t=0$ into $H(1)$ at the time $t=T$,
the Adiabatic Theorem states that the time evolution operator $U(T)$
possesses the property
\begin{equation}
    \label{eqn:ad:theorem}
    U(T) \ket{\Psi_j(0)} = \ket{\Psi_j(1)} + \Order{\left(\frac{1}{T}\right)}
\end{equation}
One prerequisite is that
there must not be any level crossings in the interval $0 \leq s \leq 1$
i.e. $E_j(s) \neq E_k(s)$ for all $k$ other than $j$.

In the limit $T \to \infty$, the final state of the evolution is
therefore the ground state of $H(1)$, provided that the
starting state was the ground state of $H(0)$.
In reality, however, the duration~$T$ of the time evolution is finite
and the question that has to be addressed is how large the duration~$T$ must be,
such that, with a high probability, the system stays in the
ground state. In other words, the probability that eigenstates
with higher energy are excited must be negligible.
A rough criterion on the duration T that guarantees that
the excitation probability is negligible reads
\begin{displaymath}
T \gg \frac{\Varepsilon}{\Delta^2}.
\end{displaymath}
$\Delta$ thereby denotes the minimum-energy difference between
the ground state and the first excited state of $H(s)$. $\Varepsilon$ corresponds
to the amplitude of the transition driven by the perturbation
$\partial H(s)/ \partial s$ between the ground state and the first excited state.
$\Varepsilon$ scales polynomially with the number of particles,
such that the duration mainly depends on the behavior of
the minimum-energy difference $\Delta$ as a function of the particle
number $n$. The energy difference usually reaches its
minimum at an avoided level crossing. At an avoided level
crossing, the ground state and the first excited state approach
as the number of particles increases, such that the duration~$T$
required for adiabicity will always increase as the number of particles
increases.

The knowledge of the spectrum of the Hamiltonian $H(s)$ is therefore necessary to make statements about the duration of the evolution.
However, the spectrum is not known, in general.
Because of this difficulty, it is advantageous to
use a simple experimental
method to check whether a
chosen evolution time~$T$ is sufficient or not~\cite{murg04}:
first, the system is
prepared in the ground state of the beginning Hamiltonian~$H(0)$.
Then, the parameter~$s$ is increased with a chosen time~$T$ up to~$1$
and decreased again with the same change rate to~$0$.
At the end, a measurement in the eigenbasis of the beginning
Hamiltonian is performed (which is known). From
this measurement it can be deduced whether the system
is still in the ground state or whether levels with
higher energy have been excited. If the system is
still in the ground state, the time evolution was adiabatic and
the evolution time~$T$ was large enough. Otherwise, the time~$T$
must be increased and the experimental check must be performed
once again.

In this paper, we focus on the XY model in a magnetic field
with Jordan-Wigner (JW) and open boundary conditions.
As mentioned before, this model depends on three parameters: the exchange energy~$J$,
the anisotropy~$\delta$ and the magnetic field strength~$B$
[see Eq. (\ref{eq:Hamiltonian simplified})].
The parameter that is varied during the evolution is $J$,
i.e. $H=H(J)$ and $J$ is ramped linearly from $0$ to $\Jmax$
during time~$T$.

In Appendix~\ref{ap:diagonalization} we review the exact diagonalization of the XY and the Ising--Hamiltonian with JW boundary conditions and present the numerically obtained spectrum of those Hamiltonian with open boundary conditions.
For $B>0$, $H(J)$ possesses level crossings between the ground state
and excited states caused by the symmetries of the Hamiltonian:
parity and, in case of JW boundary conditions, momentum.
The level crossings take place between states of different symmetry (see Appendix A).
They will not be relevant in our case,
because we investigate the spectrum with an adiabatic evolution that
is restricted to the subspace with the same symmetries
as the starting state.
The subspaces with different symmetries can therefore be treated separately:
starting with the eigenstate of $H(0)$ with parity~$p$ and momentum~$k$
that has lowest energy,
the ground state of the Hamiltonian $H(J)$ projected on the subspace
$\mathcal{H}(p,k)$ spanned by states with parity~$p$ and momentum~$k$
can be investigated.
By performing the simulation for all values of $p$ and $k$ and
comparing the respective energies, the true ground state
can be identified.
In this way, it is possible to investigate the ground state of the XY model
as a function of~$J$.

\emph{Quantum quenching} has been studied in several quantum systems that exhibit a quantum phase transition. It is a rapid drive of the system through the critical point by changing some parameter of the Hamiltonian. In this sense it is very similar to what is done in the adiabatic evolution, with the difference that, here, the relevant parameter is changed intensionally fast. The effect of the quantum quenching can be seen when the system is originally prepared in the ground state and afterwards driven trough a critical point. Since the evolution is no longer adiabatic, the system does not follow the ground state anymore, but gets excited instead. To describe those excitations in the context of thermodynamical phase transitions, the so--called \emph{Kibble-Zurek mechanism} \cite{Kibble-KZMmechanism,Zurek-KZMmechanism} was introduced. Due to the non--adiabatic evolution, defects or {\it kinks} are generated. The Kibble--Zurek mechanism predicts a certain dependency of the number of kinks on the quench time, $\tau_Q$. This dependency has been tested and verified in several experiments (see for instance \cite{ChEr12} and references therein).
In \cite{ZuDo06} those investigations have been generalized to quantum phase transitions. In particular, it has been shown there that the density of kinks, which occur in the 1D quantum Ising chain scales as $\tau_Q^{-1/2}$. In Sec \ref{sec:quenching} we will derive a matchgate circuit, which can be used to measure this relation between the number of kinks, and the quenching time for the ID Ising model. Then, we will show that this circuit can be compressed into a quantum circuit of exponentially smaller width.

Let us now derive the time evolution operators corresponding to the adiabatic as well as the non--adiabatic evolution. Consider the time dependent Hamiltonian $H(t)=\sum_{k=1}^{K}{H^{(k)}(t)}$, where $H^{(k)}$ acts on a constant number of qubits and $||H^{(k)}(t)||$ is bounded by a constant. Here, $||A||$ denotes the operator norm for an operator $A$. The Schr\"odinger equation $\dot{U}(0,T)=-iH(T) U(0,T)$ is governing the evolution operator $U(0,T)$ from time 0 to time $T$. Denoting by ${\cal T}$ the time ordering operator, its solution is given by 
\be
	U(0,t)={\cal T}(e^{-i\int_0^T H(s) ds}).
\ee

If $H(t)$ varies slowly enough in time, the time evolution operator can be evaluated as follows. One divides the total evolution time $T$, into $L+1$ steps, where the step size is chosen such that the Hamiltonian can be assumed to be constant in the interval $[t_{l-1},t_l]$, with $t_l=T/(L+1) l \equiv\Delta t l $ and $l=0,\ldots L$. In the adiabatic limit, i.e. for $T,L \rightarrow\infty$ and $\Delta t \rightarrow 0$, the time evolution operator is given by
\bea U(0,T)=\prod_{l=0}^L e^{-i H(t_l) \Delta t}.\eea In order to derive then a decomposition into elementary gates, the Trotter expansion,
\be
	e^{-i\sum_{j=0}^{M}{A_j \Delta t}}=\prod_{j=0}^{M}{e^{-i A_j \Delta t}}+\Order{(M\Delta t^2)}.
\ee
 for arbitrary operators $A_j$, is used to obtain
 \bea \label{adiab} U(0,T)=\prod_{l=0}^L \prod_{k} e^{-i H^{(k)}(t_l) \Delta t}+{\cal O} (K\Delta t^2).\eea

In the case of the XY--model we assume that the exchange energy $J$ is varied linearly in time, i.e. $J_l\equiv J(l\Delta t)=\Jmax l/L$. Then, $H(t_l)\equiv H_l$ is given by
\be
	H_{l}=-B H_0 -J_{l}(H_1+\delta H_2),
	\label{eq:Hl}
\ee
for $0 \leq l \leq L$. Let us denote by $\tilde{U}(J)$, the time evolution operator $U(0,T)$, where $T=(L+1) \Delta t$. In the adiabatic limit $\tilde{U}(J)$ is then given by
\be
	\tilde{U}(J)=\prod_{l=0}^{L} {\tilde{U}_l},
	\label{eq:digital adiabatic unitary}
\ee
where $\tilde{U}_l=e^{-i H_l \Delta t}$, with $\Delta t= T/(L+1)$. We approximate now each factor $\tilde{U}_l$ using the Trotter expansion,
\be
	\tilde{U}_l=U_0(\omega_0) U_1[\omega_1(l)] U_2[\omega_2(l)] +\mathcal{O}({\Delta t}^2),
	\label{eq:Troterization}
\ee
where $U_j(\omega_j)=e^{-i \omega_j H_j/2}$, for $j=0,1,2$ and $\omega_0=-2B\Delta t$, $\omega_1(l)=-2J_l\Delta t$ and $\omega_2(l)=-2J_l\delta \Delta t$. Thus, the unitary
\be
	U(J)=\prod_{l=0}^{L}{U_0(\omega_0) U_1[\omega_1(l)] U_2[\omega_2(l)]},
	\label{eq:U=U0U1U2}
\ee
approximates $\tilde{U}(J)$ up to $\mathcal{O}{ \left( L K {\Delta t}^2 \right) }$.
Due to the adiabatic theorem, $U(J)$ transforms the ground state of $H_0$, $\ket{\Psi(0)}$ into the ground state, $\ket{\Psi(J)}$, of $H(J)$.

Considering non--adiabatic evolutions the assumption that the Hamiltonian is constant in the interval $[t_{l-1},t_l]$ is no longer valid. Thus, even though the time evolution operator, $U(t_l,t_{l+1})$  for a time step from $t_l$ to $t_{l+1}$ can still be approximated by
\bea
	U(t_l,t_{l+1})\approx \prod_k {\cal T} (e^{-i\int_{t_l}^{t_{l+1}} H_k(s) ds}), 
\eea
with an error of $\Order(K \Delta t^2)$, the time--ordering operator has to be taken into account. However, in \cite{Ver11} it has been shown that the time--ordered integral can be approximated by the integral up to second order in $\Delta t$. More precisely, it has been shown that
\be
	\ba
		&||{\cal T} (e^{-i\int_{t_l}^{t_{l+1}} H^{(k)}(s) ds})- (e^{-i\int_{t_l}^{t_{l+1}} H^{(k)}(s) ds})||\leq \\
		&\leq \frac{2}{3} ||H^{(k)}||^2 \Delta t^2,
	\ea 
	\label{Eq:approxH}
\ee
where $||H^{(k)}||=\mbox{sup}_{0\leq s\leq 1} ||H^{(k)}(s)||$.

In case of the XY--model the (not time--ordered) integrals can be easily performed as we will see below. Let us note here that in \cite{Ver11} it has been shown that the operators $\exp(-i\int_{t_l}^{t_{l+1}} H^{(k)}(s) ds)$ can be easily approximated by a product of unitary gates of the form $\exp(-i H^{(k)}(\tau_i))$ not involving any integral. In particular, it has been proven that the time averaged Hamiltonian can be approximated by a finitely many terms $H^{(k)}(\tau_i)$, where $\tau_i$ is chosen randomly within the considered time interval. This fact led to a decomposition of the time evolution operator into polynomially many (in case $K$ is a polynomial of the number of constituting subsystems) unitary gates, just as in Eq. (\ref{adiab}).

However, for our purpose, it is enough to use Eq. (\ref{Eq:approxH}). Let us consider now the case where $B$ is varied as a function of time.
In particular, we consider $B(t)=\Bmax (1- t/T)$. Thus, we have that the terms $\exp(-i\int_{t_l}^{t_{l+1}} H^{(k)}(s) ds)$ are given by 
\bea e^{i B_l H_0 \Delta t }+{\cal O}(\Delta t^2/T), e^{i J H_1 \Delta t }, e^{i J \delta   H_2\Delta t}, \eea
where $B_l=\Bmax(1-l/L)$. Using then the Trotter formula we obtain for the evolution operator describing the quenching:

\be
	U_Q(J)=\prod_{l=0}^{L}{U_0[\omega^Q_0(l)] U_1(\omega^Q_1) U_2(\omega^Q_2)},
	\label{eq:UQ=U0U1U2}
\ee
with an error of $\Order(L \Delta t^2)$, where $U_i$, for $i=0,1,2$ is given below Eq. (\ref{eq:digital adiabatic unitary}), and $\omega^Q_0(l)=-2B_l\Delta t$, $\omega^Q_1=-2J\Delta t$ and $\omega_2=-2J\delta \Delta t$.

\section{Magnetization of the 1D XY model} \label{sec:magnetization}

The quantum phase transition can be observed by measuring the magnetization as a function of the ratio between the exchange energy $J$, and the strength of the external magnetic field $B$ [see Eq. (\ref{eq:Hamiltonian simplified})]. As mentioned above, in second order phase transitions, the non analyticity of the ground state energy only occurs if the number of spins tends to infinity. However, for a large enough system size the abrupt behavior near the critical point can be observed.

In ref. \cite{Kraus-matchgates} one of us derived a matchgate circuit, which can be utilized to measure the magnetization of the $n$-qubit Ising model (with open boundary conditions) as a function of $J$, displaying the quantum phase transition for sufficiently large $n$. Moreover, it has been demonstrated there, how the whole matchgate circuit can be compressed into a universal quantum circuit, running only on $\log{(n)}+1$ qubits. Using the symmetry of the Ising model, it has then been proven that this circuit can be further compressed into one processing only $\log{(n)}$ qubits. In this section we extend this procedure to the XY--model, for both, open and JW boundary conditions. Then we show that the compressed quantum circuit does not only utilize an exponential smaller number of qubits, but also that its size is much smaller than the one of the original matchgate circuit.

\subsection{Matchgate circuit construction}\label{sec:magnetization-matchgate}

\begin{figure}[h]
  \centering
  \includegraphics[width=0.23\textwidth]{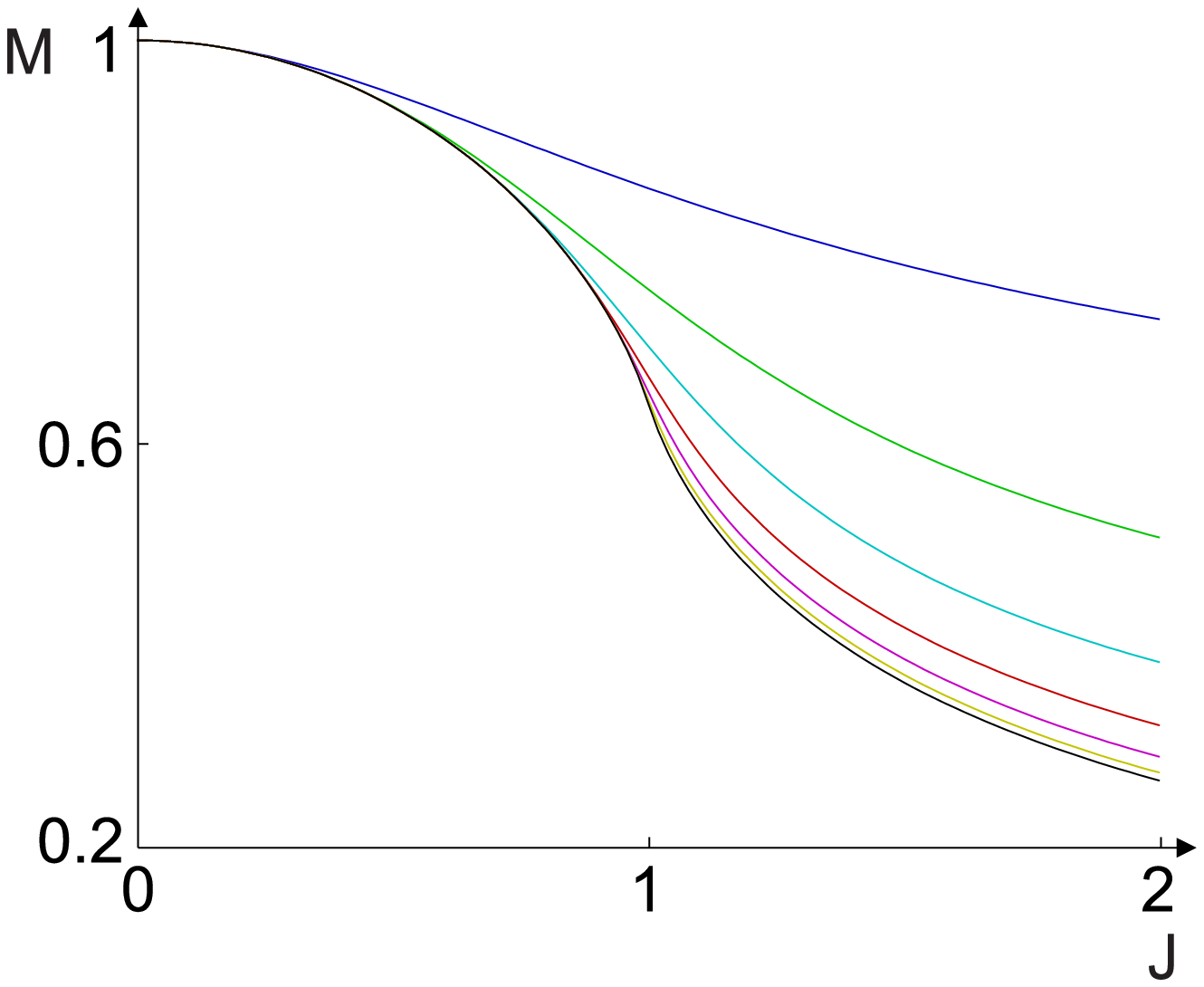}
  \includegraphics[width=0.23\textwidth]{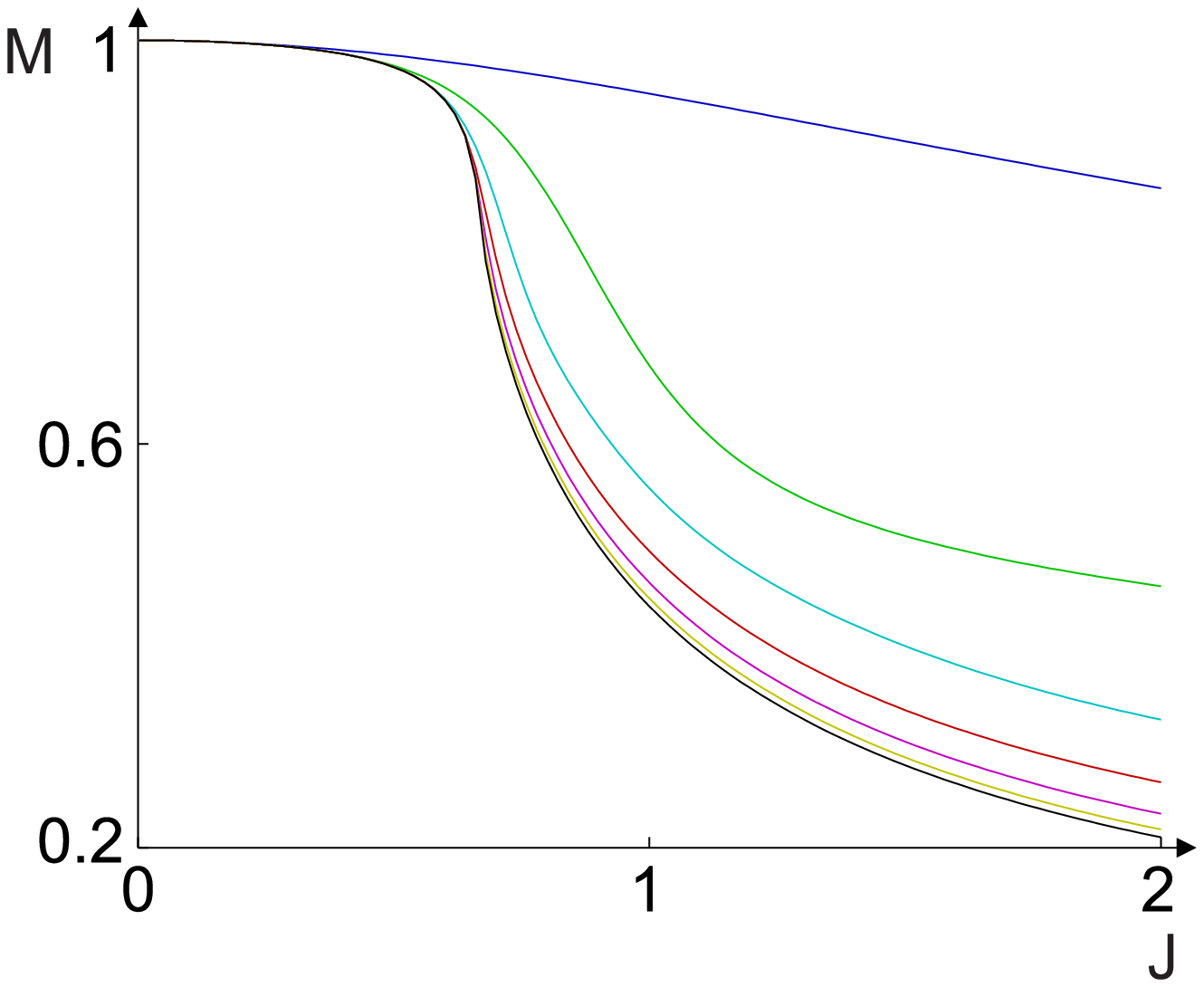}
  \caption{Magnetization of the XY Hamiltonian with JW boundary conditions, for different system sizes ($n=4,8,16,32,64,128,256$ from top to bottom respectively) and for $\delta=0$ (left) and  $\delta=0.5$ (right). This is the magnetization that will be measured when evolving the system by adiabatic evolution starting from the ground state at $J=0$ [see also Appendix \ref{ap:diagonalization}].}
  \label{fig:magnetization-adiabaticevolution}
\end{figure}

Similar to \cite{Kraus-matchgates} we construct a matchgate circuit whose output is the magnetization as a function of $J$, keeping $B$ and $\delta$ constant. We write the Hamiltonian given in Eq. (\ref{eq:Hamiltonian simplified}) as $H(J)$. The magnetization is given by 
\be 
	M(J)=\frac{1}{n}\sum_{k=1}^{n} \bmk{\Psi(J)}{Z_k}{\Psi(J)},
\ee 
where $\ket{\Psi(J)}$ denotes the ground state of $H(J)$. As can be seen in Fig. \ref{fig:magnetization-adiabaticevolution}, if the system size is large enough, the quantum phase transition can be clearly seen by measuring the magnetization as a function of $J$.

In order to measure $M(J)$, one can prepare the system in the state $\ket{\Psiin}=\ket{0}^{\otimes{n}}$, which is the ground state of $H(0)$. Then, the parameter $J$ is changed adiabatically. This evolution can be simulated digitally, as we explained in section \ref{sec:Adiabatic} by applying $U(J)$ in Eq (\ref{eq:U=U0U1U2}). In Appendix \ref{ap:diagonalization} we show that several level crossings occur in the spectrum of $H(J)$ (JW boundary conditions). However, as we explain there, the ground state of $H(0)$ has even parity and {\it momentum} (another conserved quantity) zero, which ensures that those level--crossings do not affect the adiabatic evolution. 

A simple decomposition of $U(J)$ into matchgates is obtained as follows. The unitaries $U_0$, $U_1$ and $U_2$ can be written as a product of single or two-qubit unitaries as $U_k(\omega)=\prod_{j} V_k^{(j)}(\omega)$, for $k=0,1,2$ where $V_0^{(j)}(\omega)=e^{-i \omega Z_j /2},$ $V_1^{(j)}(\omega)=e^{-i\omega X_j X_{j+1} /2},$ and $V_2^{(j)}(\omega)=e^{-i \omega Y_j Y_{j+1} /2}$. Replacing them in Eq. (\ref{eq:U=U0U1U2}) leads to
\be
	U(J)=\prod_{l=0}^{L(J)}\left[ \prod_{j}{V_0^{(j)}(\omega_0)}.\prod_{k=1}^{2}\prod_{j}{V_k^{(j)}[\omega_k(l)]}\right],
	\label{eq:U as product of Vk}
\ee
where in practice, we adjust the number of Trotter steps according to $L(J)=L J/{\Jmax}$, instead of considering a constant value $L$, to avoid performing unnecessary steps for low values of $J$. 
The second product over $j$ in Eq.(\ref{eq:U as product of Vk}) is performed for $j$ ranging between $1$ to $n$ or $n-1$, for JW or open boundary conditions respectively. It can be easily seen that $V_1^{(j)}$ and $V_2^{(j)}$ are of the form of Eq. (\ref{eq:Matchgate}) and are therefore matchgates. Similarly,  $V_0^{(j)}$ can be written as in  Eq. (\ref{eq:Matchgate}) by considering it as a gate acting as a single qubit gate on qubit $j$ and trivially on qubit $j+1$.

Since the unitary operator $U(J)$ approximates $\tilde{U}(J)$ in Eq. (\ref{eq:digital adiabatic unitary}), the state $\ket{\Psi(J)}=U(J)\ket{\Psiin}$ approximates the ground state of $H(J)$. The magnetization as a function of $J$ [up to $\mathcal{O}{ \left( L {\Delta t}^2 \right) }$] is thus given by
\be
	\ba
		M(J)=\frac{1}{n}\sum_{k=1}^{n}{\bmk{\Psiin}{U^{\dagger}(J) Z_k U(J)}{\Psiin}}.
	\ea
	\label{eq:Magnetization}
\ee
From the last expression it is evident that the following matchgate circuit can be employed to measure the magnetization as a function of $J$:
\begin{enumerate}[(i)]
	\item	Prepare the initial state $\ket{\Psiin}=\ket{0}^{\otimes n}$, i.e. the ground state of  $H(0)=-B H_0$;
	\item	evolve the system adiabatically to $\ket{\Psi(J)}$, by applying the unitary $U(J)$ in Eq. (\ref{eq:U as product of Vk}), for a certain value of $J$;
	\item	measure the $k$-th qubit in the $z$ direction to obtain the expectation value $\bk{Z_k(J)}$;
	\item	repeat the previous step for all $k$ to compute the magnetization $M(J)=\frac{1}{n}\sum_{k=1}^{n}{\bk{Z_k(J)}}$;
	\item	repeat the whole procedure for different values of $J$ between $0$ and some $\Jmax$ to obtain the magnetization as a function of $J$.
\end{enumerate}

In Fig. \ref{fig:magnetization-adiabaticevolution}, the magnetization $M(J)$ for JW boundary conditions for different values of $\delta$ and $n$, which could be measured using the circuit described above, is depicted.

\subsection{Construction of the compressed quantum circuit}\label{sec:magnetization-compressed}
Since the whole circuit presented above is a matchgate circuit it can be compressed into a quantum circuit processing only ${\cal O}[\log(n)]$ qubits. We derive here this compressed quantum circuit. More precisely, we derive a circuit running on $\log{(n)}+1$ qubits, which simulates exactly the original matchgate circuit presented above. After that, we use the symmetry of the XY--model to show that this circuit can be compressed even further to one of width $\log{(n)}$. Finally, we show that the size of the compressed circuit is drastically smaller than the one of the original matchgate circuit.

The main idea to construct the compressed circuit is the following. Let us denote by $R(J)$ the $2n \times 2n$ real orthogonal matrix corresponding to the matchgate circuit $U(J)$ given in Eq. (\ref{eq:U=U0U1U2}) (see Eq. (\ref{eq:UcU})). We will compute $R(J)$ below. As we have seen before, the final expectation value of $Z_k$, for any $k$ can be obtained by applying $U(J)$ to the initial $n$--qubit state $\ket{0}^{\otimes n}$. However, as we will show below, it can also be obtained by applying $R(J)$ to a properly chosen $\log(n)+1$--qubit input state. The output of the circuit, which is again the expectation value of a single qubit measurement coincides with $M(J)$.

We denote the basis elements of a $2n$--dimensional Hilbert space as $\ket{k}$ for $1\leq k\leq 2n$. In order to view this Hilbert space as the one corresponding to $\log(n)+1$ qubits, we use the binary notation and write $\ket{k}=\ket{k_1 k_2\dots k_m}$, where  $m=\log{(n)}+1$ and $k=\sum_{l=1}^{m-l} 2^l k_l +1$. Note that we can interpret the matrices $R$ and $S$ in Eq. (\ref{eq:Expectation Zk2}) and (\ref{eq:S}) as operators acting on a system of $m$ qubits, e.g. $S=iY_m$ acts non--trivially on qubit $m$. Since $\bk{Z_k}=\bmk{2k-1}{RSR^T}{2k}$ [see Eq. (\ref{eq:Expectation Zk2})] the magnetization is given by
\be
	M(J)=\frac{1}{n}\sum_{k=1}^n{\bmk{2k-1}{R(J)S R(J)^{T}}{2k}}.
	\label{eq:Magnetization as projector}	
\ee
Using that $S$ can be written as  $S=iY_m=\sum_{k=1}^n \kb{2k-1}{2k}-\kb{2k}{2k-1}$ and the fact that the operator $R(J)S R(J)^{T}$ is antisymmetric one can rewrite Eq. (\ref{eq:Magnetization as projector}) as
\be
	\ba
		M(J)=\frac{1}{2n}\tr{\left[R(J) Y_m R(J)^{T} Y_m\right]}.
	\ea
	\label{eq:Magnetization function of R}
\ee
The last expression can be written as the expectation value of $Y_m$ by using that $R$ is orthogonal,
\be
	M(J)=\tr{[R(J)\rhoin R(J)^T Y_m]},
	\label{eq:M(rho)}
\ee
where the $m$-qubit mixed state $\rhoin=\frac{1}{2n}(Y_m +\one)$, which can also be written as
\be
	\rhoin=\frac{\one}{n}\otimes \ket{+_y}_m\bra{+_y}_m.
	\label{eq:rhoin}
\ee

Due to Eq. (\ref{eq:M(rho)}) and the fact that $R(J)$ is real, the following circuit, which is acting on $\log(n)+1$ qubits, outputs the magnetization as a function of $J$:

\begin{enumerate}[(i)]
	\item	Prepare the initial state $\rhoin$ in Eq. (\ref{eq:rhoin});
	\item	evolve the system by the action of $R(J)$ [computed below, see Eq. (\ref{eq:R=R0R1R2})],  for a certain value of $J$;
	\item	measure the observable $Y_m$ to obtain the magnetization $M(J)=\tr{[R(J)\rhoin R(J)^T Y_m]}$;
	\item 	repeat the whole procedure for different values of $J$ between $0$ and some $\Jmax$ to obtain the magnetization as a function of $J$.
\end{enumerate}

\begin{figure}[h]
  \centering
  \includegraphics[width=0.5\textwidth]{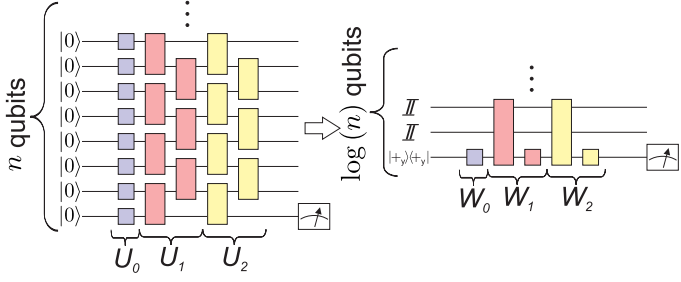}
  \caption{Schematic representation of one Trotter step of the circuits proposed to measure the magnetization of the XY model Hamiltonian. In the left side, we represent the circuit given in Sec. \ref{sec:magnetization-matchgate}. The unitaries $U_k$ for $k=0,1,2$ are given right after Eq. (\ref{eq:Troterization}), while the decomposition into single and two qubit gates is shown in Eq. (\ref{eq:U as product of Vk}). In the right side we represent the compressed quantum circuit  given in Sec. \ref{sec:magnetization-compressed2}, where the unitaries $W_k$ for $k=0,1,2$ are presented in Eq. (\ref{eq:W0,W1,W2}).}
  \label{fig:schematic circuit}
\end{figure}

It is important to note here that the compressed circuit is indeed a simulation of the original one (see Fig. \ref{fig:schematic circuit}). That is, applying the unitary $U(J)$ for some value of $J$, in the original matchgate circuit amounts to applying the unitary $R(J)$ for the same value of $J$ in the compressed circuit. Moreover, the errors due to the Trotter expansion coincide. Thus, the realization of the compressed circuit can not only be used to obtain $M(J)$, but also to measure other quantities, such as correlations, e.g. $X_j X_{j+1}$, for any $j$.

\subsubsection*{Derivation of the matrix $R(J)$}
We compute here the matrix $R(J)$ in Eq. (\ref{eq:M(rho)}).
To this end, we have to derive the real orthogonal matrices, $R_j$, associated to unitaries of the form
\be
	U_j(\omega_j)=e^{-i\omega_j H_j/2},
\ee
for $j=0,1,2$, [see Eq. (\ref{eq:H0,H1,H2})]. Here, the hat superscripts will be added to $H_j$ and $U_j$ when JW boundary conditions are considered.

First of all, we write the Hamiltonians $H_j$ and $\widehat{H}_{j}$, with $j=0,1,2$ as in Eq. (\ref{eq:Quadratic Hamiltonian}) by considering the representation of Eq. (\ref{eq:Clifford algebra}) and obtain
\be
	\ba
		H_0&=\sum_{j=1}^{n}{Z_j}=-i\sum_{j=1}^{n}{c_{2j-1}c_{2j}}, \\
		H_1&=\sum_{j=1}^{n-1}{X_jX_{j+1}}=-i\sum_{j=1}^{n-1}{c_{2j}c_{2j+1}}, \\
		H_2&=\sum_{j=1}^{n-1}{Y_jY_{j+1}}=i\sum_{j=1}^{n-1}{c_{2j-1}c_{2j+2}}.
	\ea
	\label{eq:Hs}
\ee
For the JW boundary conditions. we have that $\widehat{H}_0=H_0$, $\widehat{H}_1=H_1-i{c_{2n}c_{1}}$ and $\widehat{H}_2=H_2+i{c_{2n-1}c_{2}}$. From this representations it is easy to read off the matrices $h$ (cf. Eq. \ref{eq:Quadratic Hamiltonian}). For instance, we have
\be
	h_1=-\frac{1}{2}\sum_{j=1}^{n-1}{\kb{2j}{2j+1}}-h.c.
	\label{eq:h1}
\ee

As mentioned before, the corresponding rotation matrices $R_j$ are then given by $R_j=e^{2\omega_j h_j}$, (see Eq. (\ref{eq:UcU})). Thus, we have that
\be
	\ba
		R_0(\omega_0)&=\cos{(\omega_0)}\one + 2\sin{(\omega_0)}h_0,\\
		R_1(\omega_1)&= \cos{(\omega_1)}\one + 2\sin{(\omega_1)}h_1+\\
			&\quad \left[1-\cos{(\omega_1)}\right] (\kb{1}{1}+\kb{2n}{2n}),\\
		R_2(\omega_2)&=\cos{(\omega_2)}\one + 2\sin{(\omega_2)}h_2+\\
			&\quad \left[1-\cos{(\omega_2)}\right] (\kb{2}{2}+\kb{2n-1}{2n-1}) \\
			&= X_m R_1^T (\omega_2) X_m,
	\ea
	\label{eq:R0R1R2}
\ee
while for the JW boundary conditions the matrices are given by
\be
		\widehat{R}_j(\omega_j)=\cos{(\omega_j)}\one + 2\sin{(\omega_j)}\widehat{h}_j,
		\label{eq:R0R1R2JW}
\ee
for $j=0,1,2$. Due to Eq. (\ref{eq:U=U0U1U2}) and Eq. (\ref{eq:R0R1R2}) the matrix $R(J)$ associated to $U(J)$ is thus given by
\be
	R(J)=\prod_{l=0}^{L(J)}{R_0(\omega_0) R_1[\omega_1(l)] R_2[\omega_2(l)]},
	\label{eq:R=R0R1R2}
\ee
with $\omega_0=-2B\Delta t$, $\omega_1(l)=-2J_l \Delta t$ and $\omega_2(l)=-2J_l \delta\Delta t$ and similarly for the JW boundary conditions.

\subsection{Further compression to $\log{(n)}$ qubits}\label{sec:magnetization-compressed2}

In \cite{Kraus-matchgates} it was shown that the symmetries of the Ising model allows to construct a quantum circuit that runs only on $\hat{m}=m-1=\log (n)$ qubits. We show now that for the XY model the same procedure works, for both, open and JW boundary conditions.

The idea was to find a unitary operator, $V$, such that all the matrices $R$, $Y_m$ and $\rhoin$ in Eq. (\ref{eq:M(rho)}) transform under conjugation by $V$ into block-diagonal matrices of the form $\kb{0}{0}\otimes O_1 +\kb{1}{1}\otimes O_2$, where $O_1$ and $O_2$ are $n\times n$ matrices. The trace in Eq. (\ref{eq:M(rho)}) can then be split into two terms, both involving traces of operators acting on $\hat{m}$ qubits. It can then be shown that those two terms coincide, which implies that the magnetization can be measured with a quantum circuit of width $\log(n)$.

The $2n\times 2n$ unitary operator
\be
	V=\frac{1}{\sqrt{2}}\sum_{k=1}^{2n}{\alpha_k \kb{k}{k}+\beta_k\kb{2n-k+1}{k}},
	\label{eq:V}
\ee
with
\be
	\ba
		\alpha_k=	\begin{cases}
						{(-1)}^{k+1} & \forall k\leq n \\
						-i 			& \forall k > n \\
					\end{cases},
		&\quad \beta_k=	\begin{cases}
						{(-1)}^{k} & \forall k\leq n \\
						-i			& \forall k > n \\
					\end{cases}
	\ea
\ee
accomplishes this task for $R_0,R_1$ (i.e. the operators occurring in the Ising model). It is straightforward to see that $V$ also transforms the remaining operators into direct sums. In order to present the transformed matrices, we denote by $R_j^{\hat{m}}$ for $j=0,1,2$ the matrices which have the same form as $R_j$ but act only on $\hat{m}$ qubits. Using the notation $\tilde{O}=V^{\dagger}OV$ for an arbitrary operator $O$ we find that
\be
	\ba
		\tilde{Y}_m&=-\kb{0}{0}\otimes Y_{\hat{m}}+\kb{1}{1}\otimes Y_{\hat{m}},\\
		\tilde{R}_j(\omega_j)&=\kb{0}{0}\otimes W_j(\omega_j)+\kb{1}{1}\otimes W_j'(\omega_k),
	\ea
	\label{eq:block diagonals}
\ee
for $j=0,1,2$. Here,
\be
	\ba
		&W_0(\omega_0)=R_0^{\hat{m}}(\omega_0)^T, \\
		&W_1(\omega_1)=T_1^{\hat{m}}(\omega_1) R_1^{\hat{m}}(\omega_1)^T, \\
		&W_2(\omega_2)=T_2^{\hat{m}}(\omega_1) R_2^{\hat{m}}(\omega_2)^T,
	\ea
	\label{eq:W0,W1,W2}
\ee
and $W_j'(\omega_j)=X^{\otimes\hat{m}}W_{j}(\omega_j)^{*} X^{\otimes\hat{m}}$ for $j=0,1,2$, where $W^{*}$ denotes the complex conjugate of $W$ in the computational basis. The unitary operators $T_1^{\hat{m}}(\omega_1)=\one+(e^{i\omega_1}-1)\ket{n}\bra{n}$ and $T_2^{\hat{m}}(\omega_1)=\one+(e^{i\omega_2}-1)\ket{n-1}\bra{n-1}$ denote phase gates. Using all that, we obtain that the operator $R(J)$ in Eq.(\ref{eq:R=R0R1R2}) transforms under conjugation by $V$ into
\be
	\tilde{R}(J)=\kb{0}{0}\otimes W(J)+\kb{1}{1}\otimes W'(J),
	\label{eq:R as function of W}
\ee
where
\be
	W(J)=\prod_{l=0}^{L(J)}{W_0(\omega_0) W_1[\omega_1(l)] W_2[\omega_2(l)]}
	\label{eq:W}
\ee
and $W'(J)=X^{\otimes\hat{m}}W(J)^{*}X^{\otimes\hat{m}}$. Inserting now the expressions (\ref{eq:block diagonals}) and (\ref{eq:R as function of W}) in Eq. (\ref{eq:Magnetization function of R}), and using the fact that $R(J)$ is real, and therefore $R(J)^T=R(J)^{\dagger}$, leads to
\be
	\ba
		M(J)=&\frac{1}{2}\tr{\left[ W(J)Y_{\hat{m}} W(J)^{\dagger}Y_{\hat{m}}\right]} \\
		+&\frac{1}{2}\tr{\left[ W'(J)Y_{\hat{m}} W'(J)^{\dagger} Y_{\hat{m}}\right]}.
	\ea
	\label{eq:M function of W}
\ee

In addition to the relationship between  $W'$(J) and $W(J)$, one can easily see that $X^{\otimes\hat{m}}Y_{\hat{m}}^{*}X^{\otimes\hat{m}}=Y_{\hat{m}}$. Moreover, the first term in Eq. (\ref{eq:M function of W}) is real, which implies that the two traces occurring in Eq. (\ref{eq:M function of W}) coincide. Finally, using the same arguments that in the previous section, we write the last expression as \footnote{Note that we used here again that the Pauli operator is traceless.}
\be
	M(J)=\tr{\left[ W(J) \rhoin^{\hat{m}} W(J)^{\dagger} Y_{\hat{m}}\right]},
	\label{eq:M function of W2}
\ee
where the initial state is now the $\hat{m}$-qubit state [cf. Eq. (\ref{eq:rhoin})]
\be
	\rhoin^{\hat{m}}=\frac{2}{n} \one_{\hat{m}-1}\otimes\kb{+_y}{+_y}_{\hat{m}}.
	\label{eq:rho_inhatm}
\ee 

Due to the Eq. ({\ref{eq:M function of W2}) we have now derived the following quantum circuit of width $\log(n)$, which outputs the magnetization as a function of $J$:
\begin{enumerate}[(i)]
	\item	Prepare the initial state $\rhoin^{\hat{m}}$ in Eq. (\ref{eq:rho_inhatm});
	\item	evolve the system by the action of the unitary $W(J)$ in Eq. (\ref{eq:W}), for a certain value of $J$;
	\item	measure the operator $Y_{\hat{m}}$ in the last qubit to obtain the magnetization, according to Eq. (\ref{eq:M function of W2});
	\item	repeat the whole procedure for different values of $J$ between $0$ and some $\Jmax$ to obtain the magnetization as a function of $J$.
\end{enumerate}

In Fig. \ref{fig:magnetization-measured}, we depict the magnetization for different values of Trotter errors (see Sec. \ref{sec:Adiabatic}), compared to its exact value.

\begin{figure}[h]
  \centering
  \includegraphics[width=0.4\textwidth]{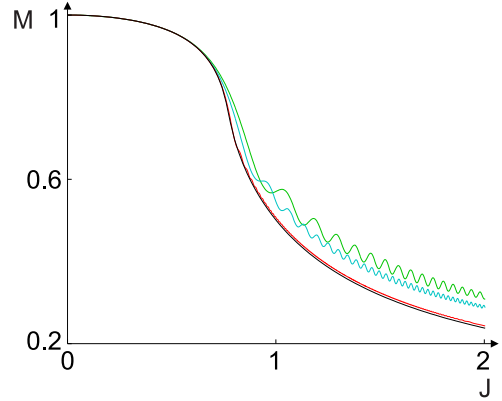}
  \caption{The magnetization obtained via the simulation of the quantum circuit of width $\hat{m}$ (see Eq. (\ref{eq:M function of W2})), compared to its exact value (black curve) for a system size of $n=128$ spins, $\delta=0.3$ and JW boundary conditions. The three curves correspond from top to bottom to the parameters of $T=50,100,1000$, while the Trotter step number is given by $L=2T^2$. Note that the Trotter error is exactly the same as in the original circuit.}
  \label{fig:magnetization-measured}
\end{figure}

\subsubsection*{Compressed circuit size}

As we have seen above, the matchgate circuit presented in Sec. \ref{sec:magnetization-matchgate} can be simulated by a circuit running on exponentially less qubits. Due to that, the latter one could be easier realizable in experiments. We show now that even the number of gates required for the realization is smaller for the compressed algorithm.
In fact, for each Trotter step, the size of the compressed circuit is exponentially smaller than the one of the matchgate circuit. Note however, that the number of required Trotter steps scales with the system size. Due to the fact that the compressed algorithm simulates the original one, the number of required Trotter steps coincide. Thus, the dominant scaling is inevitably the same for both circuits.

In order to determine the size of the compressed circuit, we decompose the matrix $W(J)$ [see Eq.(\ref{eq:W})] into single and two-qubit gates. We will show that all the matrices $W_j(\omega_j)$ for $j=0,1,2$ can be written as a product of a ${\cal O}(\hat{m})$ single qubit and controlled single-qubit gates. Using then that any $r$--qubit controlled gate can be implemented using ${\cal O}(r)$ elementary gates \cite{BarencoBenett-compression}, each Trotter step in Eq. (\ref{eq:W}) can be decomposed in  ${\cal O}(\hat{m}^2)$ elementary gates.

We consider first open boundary conditions and show then that a similar decomposition is possible for JW boundary conditions. First of all, note that
\be
	W_0(\omega_0)=O_{\hat{m}}(\omega_0),
	\label{eq:W0}
\ee
where $O_{\hat{m}}(\omega_0)$ denotes the orthogonal operator that acts non--trivially only on the $\hat{m}$-th qubit, where it acts as the single-qubit gate $O(\omega_0)=e^{i \omega_0 Y}$. To decompose $R_1^{\hat{m}}(\omega_1)$ into elementary gates, note that
\be
	R_1^{\hat{m}}(\omega_1)^T=\kb{1}{1}\oplus\left[O(\omega_1)^{\oplus (n/2-1)}\right]\oplus\kb{2n}{2n},
\ee
where $O(\omega_1)^{\oplus (n/2-1)}$ denotes the direct sum of $\frac{n}{2}-1$ identical blocks $O(\omega_1)$. In order to decompose $R_1^{\hat{m}}$ in ${\cal O}(\hat{m}^2)$ single qubit and $\hat{m}$--fold controlled unitary gates, we introduce the real orthogonal matrix
\be
	\ba
			A	&=\sum_{j=1}^{n-1}{\kb{j+1}{j}}+\kb{1}{n}\\
				&=X_{\hat{m}}\left[\Lambda^{(\hat{m})}X_{\hat{m}-1}\right]\cdots\left[\Lambda^{(\hat{m},\dots ,2)}X_{1}\right]
	\ea
	\label{eq:A}
\ee
Here, and in the following $\Lambda^{(i_1,\ldots, i_l)}O_{k}$ denotes the controlled single qubit operator $O$, acting on qubit $k$ and the qubits $i_1,\ldots i_l$ denote the controlling qubits, i.e. $\Lambda^{(i_1,\ldots, i_l)}O_{k}=(\one-\proj{1}^{\otimes l})_{i_1,\ldots i_l}\otimes \one +\proj{1}^{\otimes l}_{i_1,\ldots i_l}\otimes O_k$. Note that $A$ acts on the computational basis as $A\ket{k}=\ket{k\oplus 1}$, where $\oplus$ denotes the addition modulo $n$. It can be easily seen that
\be
	\ba
		R_1^{\hat{m}}(\omega_1)^T&= A \left\{ \left[O(\omega_1)^{\oplus (n/2-1)}\right]\oplus\one \right\}A^{T}\\
		&=A O_{\hat{m}}(\omega_1)B(\omega_1) A^T,
	\ea
	\label{eq:R1}
\ee
where $B(\omega_1)=\Lambda^{(1,\dots,\hat{m}-1)}O_{\hat{m}}^T(\omega_1)$. Using that $R_2^{\hat{m}}(\omega_2)=X_{\hat{m}} R_1^{\hat{m}}(\omega_2)^T X_{\hat{m}}$ it is straightforward to obtain a similar expression for $R_2^{\hat{m}}(\omega_2)$.

Finally, note that the matrices $T_1^{\hat{m}}$ and $T_2^{\hat{m}}$ in Eq. (\ref{eq:W0,W1,W2}) are $(\hat{m}-1)$--fold controlled single qubit gates. Together with the decompositions of the operators $W_0(\omega_0)$, $R_1^{\hat{m}}$, and $R_2^{\hat{m}}$ [see Eq. (\ref{eq:W0}) and Eq. (\ref{eq:R1})] and the fact that $A$ is decomposed into ${\cal O}(\hat{m})$ single--qubit controlled gates, this shows that every term in Eq. (\ref{eq:W}) can be implemented with ${\cal O}(\hat{m})$ single-qubit and  $(\hat{m}-1)$--fold controlled single qubit gates. As mentioned before, any $(\hat{m}-1)$--fold controlled single qubit gate can be decomposed into $\Order (\hat{m})$ elementary gates \cite{BarencoBenett-compression}. Thus, each Trotter step in the compressed circuit can be implemented with $\Order (\hat{m}^2)=\Order (\log(n)^2)$ elementary gates. In contrast to that, the original matchgate circuit presented in Sec. \ref{sec:magnetization-matchgate} requires $\poly (n)$ matchgates, as can be seen in Eq. (\ref{eq:U as product of Vk}). Hence, we have an exponential reduction in the size for each Trotter step. Note, however, as mentioned above, that the error of the adiabatic evolution presented in Eq. (\ref{eq:digital adiabatic unitary}) depends on $n$, and because of that the number of Trotter steps, $L$, scales polynomially in $n$. Hence, the size of the compressed circuit is $\Order(L\hat{m}^2)=\Order[ \mathrm{poly}(n)]$, which prevents this circuit from having an exponential gain in size over the original matchgate circuit.

In the case of JW boundary conditions a similar decomposition into ${\cal O}(\hat{m}^2)$ elementary gates can be obtained. In particular, one finds that
\be
	\ba
	\widehat{R}_1^{\hat{m}}(\omega_1)&=A^T \widehat{R}_0^{\hat{m}}(\omega_1) A\\
	\widehat{R}_2^{\hat{m}}(\omega_2)&=X_m\widehat{R}^{\hat{m}}_1(\omega_2)^T X_m.
	\ea
\ee
The unitaries $W_1(\omega_1)$ and $W_2(\omega_2)$ in Eq. (\ref{eq:W0,W1,W2}) have to be replaced by $\widehat{W}_1(\omega_1)=\widehat{T}_1^{\hat{m}}(\omega_1) \widehat{R}_1^{\hat{m}}(\omega_1)^T$ and $\widehat{W}_2(\omega_2)=\widehat{T}_2^{\hat{m}}(\omega_2) \widehat{R}_2^{\hat{m}}(\omega_2)^T$ respectively, where $\widehat{T}_1^{\hat{m}}(\omega_1)=A\Lambda^{(1,\dots,\hat{m}-1)}(C_{\hat{m}})A^{T}$ and $\widehat{T}_2^{\hat{m}}(\omega_2)=A^2\Lambda^{(1,\dots,\hat{m}-2)}(D_{\hat{m}-1,\hat{m}})(A^{T})^2$, where $C_{\hat{m}}$ acts non-trivially only on the $\hat{m}$--th qubit, where it acts as the single--qubit gate $C=e^{i\omega_1}O_{\hat{m}}^T(\omega_1)$ and $D_{\hat{m}-1,\hat{m}}$ acts on the last two qubits as the two--qubit gate $D=\one+[e^{i\omega_2}\cos(\omega_2)-1](\kb{00}{00}+\kb{11}{11})+e^{i\omega_2}\sin(\omega_2)(\kb{00}{11}-\kb{11}{00})$.

\section{Compressed circuit for quantum quenching}\label{sec:quenching}

In this section we present first a matchgate circuit to implement a quench-induced transition in the 1D Ising model. Thus, whenever we refer to a previous equation, we take $\delta=0$. Similar to \cite{ZuDo06} our aim is to measure the number of kinks as a function of the quench time. In the subsequent subsections we derive the corresponding compressed quantum circuits with width $\log(n)+1$ and $\log(n)$ respectively.

In contrast to the previous section, we are going to change here both parameters, $J$ and $B$. This is why we write the Hamiltonian in Eq. (\ref{eq:Hamiltonian simplified}) now as
\be
	H(B,J)=-B H_0-J H_1.
	\label{eq:Hamiltonian Q.Q.}
\ee

As we have seen before, if the system is prepared in the ground state of $H(B\rightarrow{\infty})$ and the parameter $B$ is decreased slowly till $B=0$, then due to the adiabatic theorem the system will evolve into a ground state of $H(B=0)$. However, if the system is quenched, i.e. the evolution is no longer adiabatic, from $B\rightarrow \infty$ to $B=0$, it evolves into some excited state. Note that while the ground state for $B\rightarrow\infty$ is $\ket{0}^{\otimes{n}}$,  for $B=0$ it is degenerate and the ground state subspace is spanned by $\ket{+}^{\otimes{n}}$ and $\ket{-}^{\otimes{n}}$. If the system is quenched from
$B\rightarrow\infty$ to $B=0$, the spin components are no longer aligned with respect to their neighboring spins, but some \emph{kinks} appear. The quantity
\be
	\nu=\frac{1}{2}(1-K),
	\label{eq:nu}
\ee
with
\be
	K=\frac{1}{n-1}\sum_{k=1}^{n-1}{\bk{X_k X_{k+1}}},
	\label{eq:K}
\ee
quantifies the number of \emph{kinks} per spin in the system. If the state evolves to the ground state, one obtains that $\nu=0$, but if the state evolves into some excited states then $\nu>0$. In the following we present a matchgate circuit and a compressed quantum circuit to measure $\nu$ as a function of the quenching time.

\subsection{ Construction of the Matchgate circuit}\label{sec:quenching-matchgate}

In the ideal case, the system is quenched from  $B\rightarrow \infty$ to $B=0$. To implement this, however, it is sufficient to start from a finite but large value of $B$ and then quench the system to $B=0$ through the critical point. The system is initially prepared in the ground state of the Hamiltonian $H(\Bmax,\Jmax)$. This can be achieved via adiabatic evolution. Since the Ising model exhibits a phase transition at the point $B\sim J$, the error induced by starting the quenching from a finite value of $B$ can be neglected as long as $\Bmax>>\Jmax$. After this adiabatic evolution the system is quenched by rapidly varying $B$ from $\Bmax$ to $0$, keeping $J=\Jmax$ constant. We denote by $U_Q^{(1)}$ and $U_Q^{(2)}$ the unitaries associated to the two evolutions respectively.

Since the first part of the evolution is done adiabatically, the matchgate circuit that simulates this evolution is given by the one presented in Sec \ref{sec:magnetization-matchgate} by considering $J=\Jmax$ in Eq. (\ref{eq:U=U0U1U2}). That is,
\be
	U_Q^{(1)}=\prod_{l=0}^{L_1(\Jmax)}{U_0(\gamma_0) U_1[\gamma_1(l)]},
	\label{eq:Ua}
\ee
where $U_j$ for $j=0,1$ are given right after Eq. (\ref{eq:Troterization}), $\gamma_0=-2B{\Delta t}_1$ and $\gamma_1(l)=-2J_l{\Delta t}_1$, where ${\Delta t}_1= T_1/(L_1+1)$.

In order to obtain $U_Q^{(2)}$ we use the same tools as before, but now considering $J$ as a constant and $B$ as a variable. The Hamiltonian is discretized into $L_2+1$ steps of the form
\be
	\tilde{H}_{l}=-B_{l} H_0-\Jmax H_{1},
	\label{eq:Hprima l}
\ee
for $0\leq l\leq L_2$, where $B_l=\Bmax(L_2-l)/L_2$. As explained in Sec. \ref{sec:Adiabatic}, the unitary describing the quantum quenching evolution is approximated (in second order) by
\be
	U_Q^{(2)}(T)=\prod_{l=0}^{L_2(\Bmax)}{U_0[\phi_0(T,l)]U_1[\phi_1(T)]},
	\label{eq:Uq}
\ee
where $\phi_0(T,l)=-2B_l{\Delta t}_2(T)$, $\phi_1(T)=-2J{\Delta t}_2(T)$ and ${\Delta t}_2(T)= T/{(L_2+1)}$. The time $T$ is the total evolution time of the quenching. The total unitary corresponding to the whole evolution is then given by
\be
	U_Q(T)=U_Q^{(2)}(T)U_Q^{(1)}.
	\label{eq:Uq2}
\ee
Hence, the number of kinks can be obtained using that
\be
	K(T)=\frac{1}{n-1}\sum_{k=1}^{n-1}{\bmk{\Psiin}{U^{\dagger}_Q(T)X_k X_{k+1}U_Q(T)}{\Psiin}}.
	\label{eq:Expectation K}
\ee
Thus, the following matchgate circuit can be used to measure the number of kinks  $\nu$ as a function of the quenching time $T$:
\begin{enumerate}[(i)]
	\item	Prepare the initial state $\ket{\Psiin}=\ket{0}^{\otimes n}$,  i.e the ground state of  $H(\Bmax,0)=-\Bmax H_0$;
	\item	evolve the system adiabatically to the ground state of $H(\Bmax,\Jmax)$, by applying the unitary $U_Q^{(1)}$ in Eq. (\ref{eq:Ua});
	\item	quench the system from $\Bmax$ to $0$ rapidly, by applying the unitary $U_Q^{(2)}(T)$ in Eq. (\ref{eq:Uq}) with a small value of $T$;
	\item	measure the two qubit correlation $X_kX_{k+1}$ to obtain $\bk{X_kX_{k+1}(T)}$;
	\item	repeat the previous step for every $k$, to compute $K(T)$ in Eq. (\ref{eq:Expectation K}) and with that the number of kinks $\nu$ in Eq. (\ref{eq:nu});
	\item	repeat the whole procedure for different values of $T$, to obtain the number of kinks as a function of the quenching time.
\end{enumerate}

\subsection{Construction of the compressed quantum circuit}\label{sec:quenching-compressed}

To derive the compressed quantum circuit to measure $K$ we proceed similarly as in Sec \ref{sec:magnetization-compressed}. Using Eq. (\ref{eq:UcU}) and the fact that $X_kX_{k+1}=-ic_{2k} c_{2k+1}$, we find for the number of kinks,
\be
	\ba
		K(T)
		&=\frac{1}{n-1}\sum_{k=1}^{n-1}{\bmk{\Psiin}{U^{\dagger}_Q(T)(-i c_{2k}c_{2k+1})U_Q(T)}{\Psiin}}	\\
		&=\frac{1}{n-1}\sum_{k=1}^{n-1}{\bmk{2k}{R_Q(T)SR_Q^T(T)}{2k+1}}\\
		&= \frac{1}{n-1}\tr{\left[R_Q(T) Y_m R_Q^T(T)(ih_1)\right]},\\
	\ea
	\label{eq:K(T)}
\ee
where we have used that $S=iY_m$. The matrix $h_1$ coincides with the one used to compute $R_1$ in Eq. (\ref{eq:R0R1R2}), and is given by Eq.(\ref{eq:h1}). The orthogonal matrix $R_Q(T)$ in Eq. (\ref{eq:K(T)}), which we compute below, is associated to the unitary $U_Q(T)$ in Eq. (\ref{eq:Uq2}).

Similarly to the previous section we rewrite Eq. (\ref{eq:K(T)}) as the outcome of a single qubit measurement. In this case, we use the fact that $\bmk{j}{R_Q(T)Y_m R_Q^T}{j}=0$ for any $j$, since $R_QY_mR_Q^T$ is antisymmetric, and thus adding the vanishing term $\frac{1}{2(n-1)}\tr{\left[R_Q(T) Y_m R_Q^T(T)\sum_{j=2}^{2n-1}{\kb{j}{j}}\right]}$ to Eq. (\ref{eq:K(T)}), one can write
\be
	K(T)=\tr{\left[R_Q^T(T) \xiin R_Q(T) Y_m \right]},
	\label{eq:K(xiin)}
\ee
where
\be
	\xiin=\frac{1}{2(n-1)}\sum_{k=1}^{n-1}{\big(\ket{2k}+i\ket{2k+1}\big)\big(\bra{2k}-i\bra{2k+1}\big)}.
\ee

Since $\xiin$ might not be easily generated in an experiment, we apply now a basis transformation to derive a more physical input state. Equivalently to $A$ in Eq. (\ref{eq:A}), we define the operator $A_m=\sum_{j=1}^{2n-1}{\kb{j+1}{j}}+\kb{1}{2n}$, which acts on $m$ qubits, and one can easily check that
\be
	\xiin= A_m\sigmain A_m^T,
	\label{eq:xi}
\ee
where
\be
	\sigmain=\frac{1}{2(n-1)}\sum_{k=1}^{n-1}{\big(\ket{2k-1}+i\ket{2k}\big)\big(\bra{2k-1}-i\bra{2k}\big)},
	\label{eq:sigmain}
\ee
which can be written as
\be
	\sigmain=\frac{\one-\kb{n}{n}}{n-1}\otimes\kb{+_{y}}{+_{y}}_m.
	\label{eq:sigmain2}
\ee

Replacing Eq. (\ref{eq:xi}) into the Eq. (\ref{eq:K(xiin)}) we find that
\be
	K(T)=\tr{\left[T_Q^T(T)\sigmain T_Q(T) Y_m \right]},
	\label{eq:K(T)final}
\ee
where $T_Q^T(T)=R_Q^T(T)A_m$. Due to Eq. (\ref{eq:K(T)final}), the following circuit, running on $\log(n)+1$ qubits, can be used to measure the number of kinks as a function of $T$:
\begin{enumerate}[(i)]
	\item	Prepare the initial state $\sigmain$ in Eq. (\ref{eq:sigmain2});
	\item	evolve the system by applying of the operator $T_Q^T(T)$;
	\item	measure the operator $Y$ in the last qubit to obtain $K(T)$ in Eq. (\ref{eq:K(T)final}) and with that the number of kinks $\nu$ in Eq. (\ref{eq:nu});
	\item	repeat the whole procedure for different values of $T$, to obtain the number of kinks as a function of the quenching time.
\end{enumerate}

\subsubsection*{Computing the matrix R}

The matrix $R_Q(T)$ can be easily constructed following the procedure of the previous section. This is due to the fact that $U_Q(T)$ is decomposed into the same gates as $U(J)$ in Eq. (\ref{eq:U=U0U1U2}). They only differ in the arguments. However, this does not change the procedure to construct $R_Q$. It can be easily seen that the matrices $R_Q^{(1)}$ and $R_Q^{(2)}(T)$ associated to the unitaries $U_Q^{(1)}$ and $U_Q^{(2)}(T)$ are given by
\be
	R_Q^{(1)}=\prod_{l=0}^{L_1(\Jmax)}{R_0(\gamma_0)R_1[\gamma_1(l)]},
	\label{eq:RQ1}
\ee
and
\be
	R_Q^{(2)}(T)=\prod_{l=0}^{L_2(\Bmax)}{R_0[\phi_0(T,l)]R_1[\phi_1(T)]}.
	\label{eq:RQ2}
\ee

Multiplying those two matrices we obtain $R_Q(T)=R_Q^{(2)}(T)R_Q^{(1)}$.

\subsection{Further compression to $\log (n)$ qubits}\label{sec:quenching-compressed2}

In order to compress the algorithm further, note that $\sigmain$ in Eq. (\ref{eq:sigmain2}) does not transform under the unitary $V$ into the desired block structure like in Eq. (\ref{eq:block diagonals}). However, $h_1$ in  Eq. (\ref{eq:h1}) does. In fact the transformed operator $\tilde{h}_1=V^{\dagger}h_1 V$ takes the form
\be
	i \tilde{h}_1=-\kb{0}{0}\otimes O+\kb{1}{1}\otimes O',
	\label{eq:tilde_h1}
\ee
where
\be
	O=\frac{1}{2}\kb{n}{n}+ih_1^{\hat{m}},
\ee
and $O'=X^{\otimes\hat{m}} O^{*} X^{\otimes\hat{m}}$.

Note that all the operators occurring in the decomposition of $R_Q$ [see Eqs. (\ref{eq:RQ1}) and (\ref{eq:RQ2})] are of the form of $R_j(\omega_j)$ in Eq. (\ref{eq:R0R1R2}) for $j=0,1,2$. As mentioned before, the only difference to the circuit which can be used to measure the magnetization lies in the coefficients $\omega_j$ but not in the structure of the matrices $R_j$. Thus, in the same way as in Eq. (\ref{eq:R as function of W}), we write
\be
	\tilde{R}_Q(T)=\kb{0}{0}\otimes W_Q(T)+\kb{1}{1} W_Q'(T),
	\label{eq:RQ as function of W}
\ee
with  $W_Q(T)=W_Q^{(2)}(T)W_Q^{(1)}$. Both factors $W_Q^{(1)}$ and $W_Q^{(2)}(T)$ are given by Eq. (\ref{eq:W}) with the arguments replaced by $\gamma_j$ and $\phi_j$ respectively for $j=0,1,2$. The upper limit $L(J)$ also has to be changed to $L_1(\Jmax)$ and $L_2(\Bmax)$ respectively.

Similarly to Sec. \ref{sec:magnetization-compressed2} we obtain, by inserting Eqs. (\ref{eq:tilde_h1}), (\ref{eq:RQ as function of W}) and (\ref{eq:block diagonals}) into Eq. (\ref{eq:K(T)}), the following expression for the number of kinks,
\be
	\ba
		K(T)&=\frac{1}{n-1}\tr{\left[ W_Q^{\dagger}(T) O W_Q(T)Y_{\hat{m}}\right]} \\
		&\quad +\frac{1}{n-1}\tr{\left[ W_Q'^{\dagger}(T) O'W_Q'(T)Y_{\hat{m}} \right]}\\
		&=\frac{2}{n-1}\tr{\left[ W_Q^{\dagger}(T) O W_Q(T) Y_{\hat{m}}\right]}.
	\ea
	\label{eq:K function of W}
\ee
where, in order to derive the second equality we used the fact that the two trace occurring in the first equality are equal, which can be easily shown.

Using the fact that $W_Q(T)$ is unitary and that $Y_{\hat{m}}$ is traceless, we can replace $-\frac{2}{n-1}O$ in Eq. (\ref{eq:K function of W}) by the density operator
\be
	\chi_{\mathrm{in}}^{\hat{m}}=\frac{1}{n-1}(\one_{\hat{m}}-2O),
\ee
which implies that
\be
	K(T)=-\tr{\left[ W_Q^{\dagger}(T)\chi_{\mathrm{in}}^{\hat{m}} W_Q(T) Y_{\hat{m}}\right]}.
	\label{eq:K function of W2}
\ee

Finally, and in the same way as we did in Sec. \ref{sec:quenching-compressed}, we apply a basis transformation $\zeta_{\mathrm{in}}^{\hat{m}}=A^T\chi_{\mathrm{in}}^{\hat{m}}A$ to derive the more physical input state
\be
	\ba
		\zeta_{\mathrm{in}}^{\hat{m}}=&\frac{1}{n-1}\Bigg[\sum_{k=1}^{\frac{n}{2}-1}{2\kb{k}{k}\otimes\ket{-_y}_{\hat{m}}\bra{-_y}_{\hat{m}}}\\
		&+\kb{\frac{n}{2}}{\frac{n}{2}}\otimes\ket{1}_{\hat{m}}\bra{1}_{\hat{m}}\Bigg].\\
	\ea
	\label{eq:sigmainhat}
\ee
In terms of this state, the Eq. (\ref{eq:K function of W2}) takes the form
\be
	K(T)=-\tr{\left[ W_Q^{\dagger}(T)A\zeta_{\mathrm{in}}^{\hat{m}} A^{T} W_Q(T) Y_{\hat{m}}\right]}.
	\label{eq:K function of sigmainhat}
\ee
Thus, $K(T)$ can be computed by the following circuit, running on $\log(n)$ qubits:
\begin{enumerate}[(i)]
	\item	Prepare the $\log(n)$--qubit initial state $\zeta_{\mathrm{in}}^{\hat{m}}$ in Eq. (\ref{eq:sigmainhat});
	\item	evolve the system by the action of the operator $W_Q^{\dagger}(T)A$;
	\item	measure the operator $Y$ in the last qubit to obtain $K(T)$ in Eq. (\ref{eq:K function of sigmainhat}) and with that the number of kinks $\nu$ in Eq. (\ref{eq:nu});
	\item	repeat the whole procedure for different values of $T$, to obtain the number of kinks as a function of the quenching time.
\end{enumerate}

In Fig. \ref{fig:kinks}, we depict the number of kinks $\nu$ as a function of $B$ [see Eq. (\ref{eq:K function of sigmainhat})], as the magnetic field is quenched from $\Bmax$ towards zero in different quenching times $T$. One can observe that as $T$ grows, the number of kinks at the end of the evolution, i.e. at $B=0$, decreases. As in \cite{ZuDo06} we observe that the density of kinks  for this value of $B$ scales like the inverse of the square root of the quenching time (see Fig. \ref{fig:linear fit}).
\begin{figure}[H]
  \centering

  \includegraphics[width=0.4\textwidth]{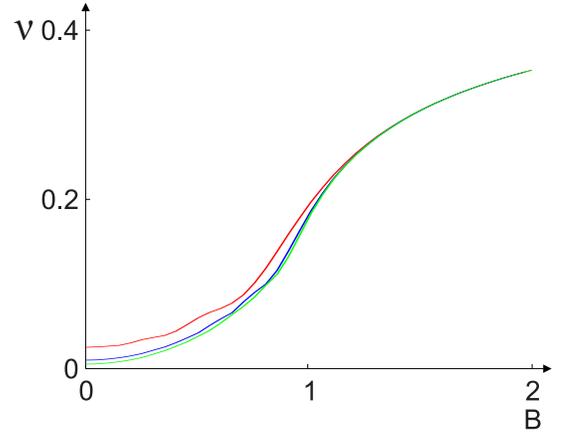}
  \caption{Number of kinks, $\nu$, as a function of $B$, during the quantum quenching evolution [see Eq.(\ref{eq:K function of sigmainhat})], for a system size of $n=128$ spins and JW boundary conditions. From top to bottom the different curves corresponds to quenching times of $T_2=50,150$ and $250$ respectively, and the number of Trotter steps, $L=2T_2^2$.}
  \label{fig:kinks}
  
\end{figure}

\begin{figure}[H]
	\centering
	\includegraphics[width=0.23\textwidth]{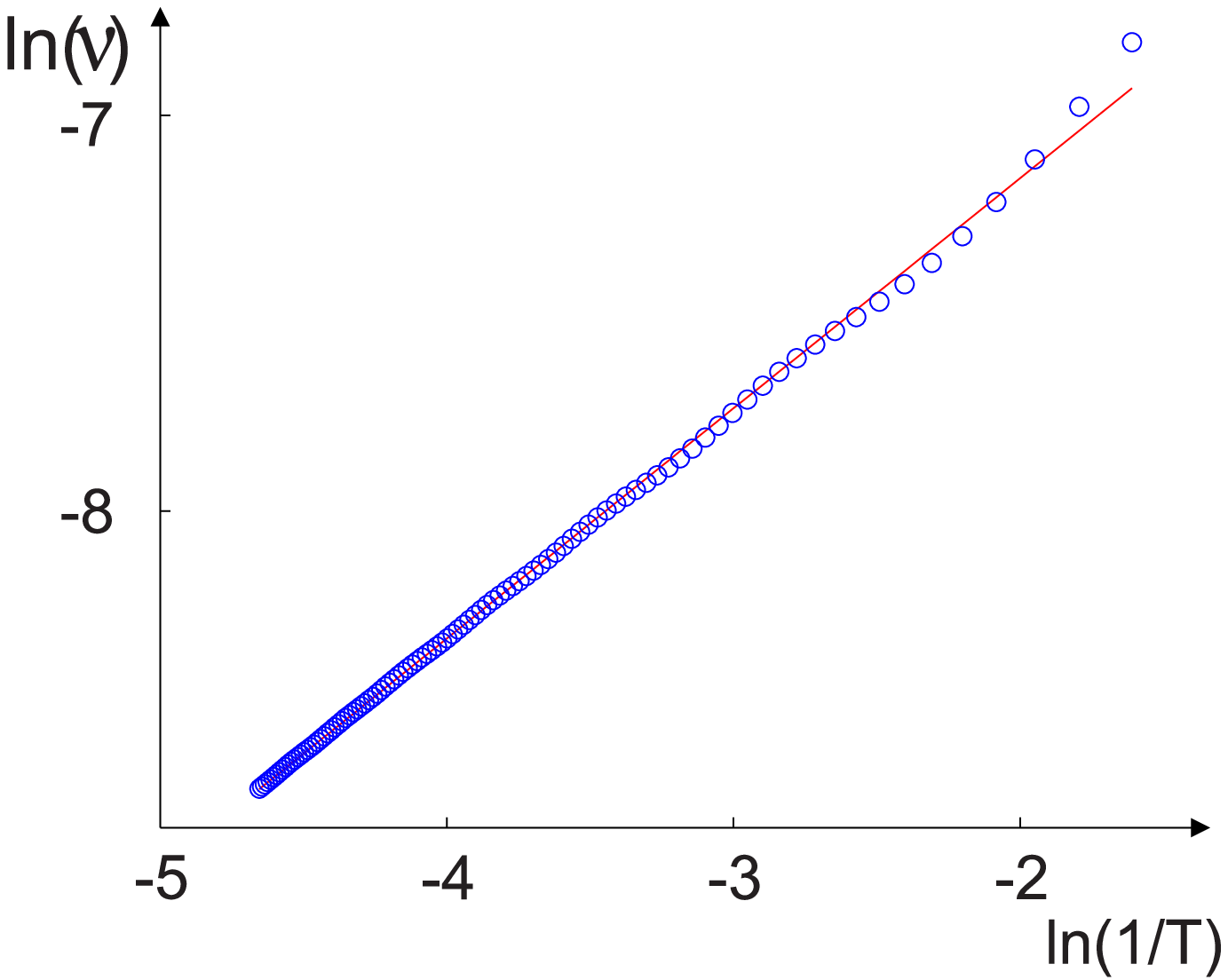}
	\includegraphics[width=0.23\textwidth]{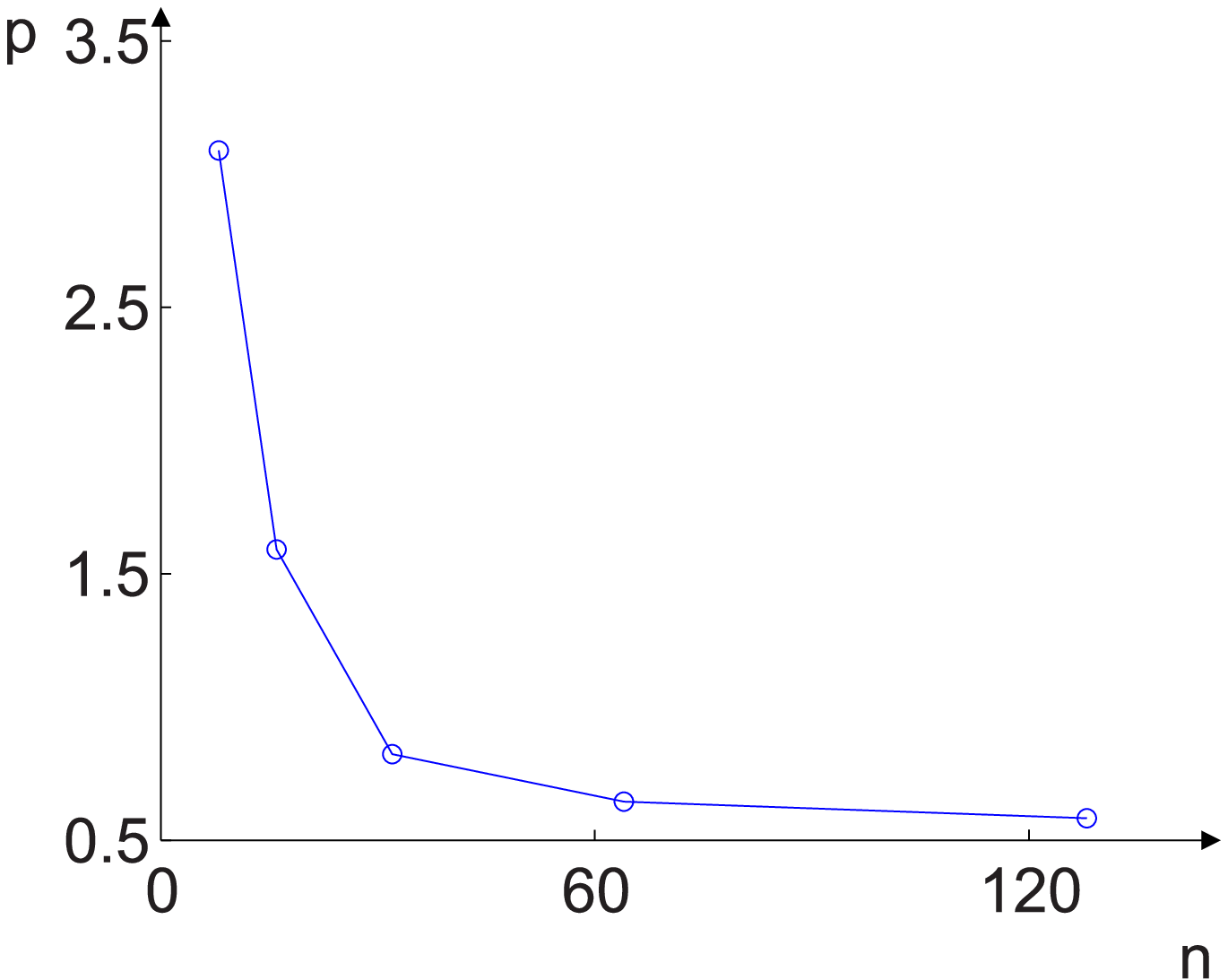}
	\caption{Logarithm of the number of kinks $\nu$ as a function of the logarithm of the inverse of the quenching time $T$. The number of kinks was obtained via the simulation of the quantum circuit of width $\hat{m}$ [see Eq. (\ref{eq:K function of sigmainhat})], at the end of the quenching evolution, i.e. at B=0 (see Fig. \ref{fig:kinks}). The system size is $n=128$. The solid line represents the fitted linear function, which gives a slope of $p=0.58$ (in agreement with \cite{ZuDo06}), that relates the number of kinks and the quenching time according to $\nu\propto T_2^{-p}$ . The value of $p$ was obtained for system sizes of $n=8,16,32,64,128$ which are plotted on the right side. Note that $p$ converges to $0.5$ with increasing system size.}
	\label{fig:linear fit}
\end{figure}

\section{Finite time evolution}\label{sec:timeevolution}

In the previous sections, we considered time dependent Hamiltonians. For example, in the adiabatic evolution or in the quantum quenching, some parameter of the Hamiltonian is either slowly or rapidly changed in time. Due to the results derived in \cite{Ver11}, also other dynamics can be studied along the same lines. Here, we consider as a last application the finite time evolution governed by a constant Hamiltonian. We study the propagation of excitations in time in a 1D spin chain. Those processes have been studied, for instance, in \cite{ZaunerNishino-ComovingWindow}. There, an infinite 1D spin chain was considered, whose evolution corresponded either to the Ising Hamiltonian, or the XYZ model Hamiltonian. Initially some of the spins were flipped and the propagation of this signal front was simulated. The time evolution was determined using matrix product states \cite{PeWo06}.

In this section we show how to construct a matchgate circuit and the equivalent compressed quantum circuit to compute the propagation of a signal in a 1D spin chain with interactions given by the XY model. The signal is generated by flipping two spins of the system, which is initially prepared in the ground state of the XY Hamiltonian. Then, the system evolves according to the XY--Hamiltonian and the spread of the signal is measured by measuring $Z_k$ for each qubit $k$. The initial preparation, the creation of the signal, the time evolution of the system, and the measurement corresponds to a matchgate circuit, as we show below.

\subsection{Constructing the matchgate circuit}\label{sec:timeevolution-matchgate}

In order to construct the ground state $\ket{\Psi(J)}$ of the Hamiltonian $H(J)$, we use the matchgate circuit introduced in Sec. \ref{sec:magnetization}. That is, the ground state of the XY Hamiltonian is obtained by adiabatically evolving the initial state $\ket{\Psiin}=\ket{0}^{\otimes{n}}$ to $\ket{\Psi(J)}$. Thus the first part of the evolution is described by the unitary
\be
	U_T^{(1)}(J)=U(J),
	\label{eq:UT1}
\ee
where $U(J)$ is given by Eq. (\ref{eq:U=U0U1U2}). To generate an excitation in the spin chain, the two spins in the middle of the chain are flipped by applying the unitary
\be
	U_T^{(2)}=iX_{n/2}X_{n/2+1}.
	\label{eq:UT2}
\ee
Clearly, this unitary is of the form of Eq. (\ref{eq:Matchgate}) and acts on nearest neighbors. Finally, to obtain the circuit that simulates the time evolution, we have to decompose the unitary
\be
	\tilde{U}_{T}^{(3)}(J,t)=e^{-iH(J)t}
	\label{eq:UT3}
\ee
into matchgates, where $H$ is the XY Hamiltonian in Eq. (\ref{eq:Hamiltonian simplified}). Defining $\Delta t_T=t/L_T$ for some integer $L_T$, one can rewrite the previous unitary as $\tilde{U}_{T}^{(3)}=\left(e^{-iH\Delta t_T} \right)^{L_T}$. The parameter $L_T$ is chosen such that $\Delta t_T$ is small. Using the Trotter formula, one finds that
\be
	\ba
		\tilde{U}_{T}^{(3)}(J,t)&=\big[U_0(\zeta_0) U_1(\zeta_1) U_2(\zeta_2)+\Order(\Delta t_T^2)\big]^{L_T}\\
			&=\big[U_0(\zeta_0) U_1(\zeta_1) U_2(\zeta_2)\big]^{L_T}+\Order(L_T \Delta t_T^2)\\
	\ea
	\label{eq:UT3=U0U1U2}
\ee
with $\zeta_0=-2B\Delta t_T$, $\zeta_1=-2J\Delta t_T$ and $\zeta_2=-2J\delta \Delta t_T$.
Thus, the unitary that approximates $\tilde{U}_T^{(3)}$ is given by
\be
	U_{T}^{(3)}(J,t)=\Bigl\{U_0\bigl[\zeta_0(t)\bigr] U_1\bigl[\zeta_1(J,t)\bigr] U_2\bigl[\zeta_2(J,t)\bigr]\Bigr\}^{L_T}.
	\label{eq:UT3fin}
\ee

The unitary that governs the whole evolution is given by
\be
	U_T(J,t)=U_{T}^{(3)}(J,t)U_{T}^{(2)}U_{T}^{(1)}(J).
	\label{eq:UT=Ut1Ut2Ut3}
\ee
The spread of the excitation as a function of time can be measured by measuring $Z_k$
for every qubit $k$. Its expectation value is given by

\be
	\bk{Z_k(J,t)}=\bmk{\Psiin}{U_T^{\dagger}(J,t)Z_kU_T(J,t)}{\Psiin}.
	\label{eq:Zk(J,t)}
\ee
 Thus, we have shown that in order to measure the propagation of a signal in the 1D chain, which was initially generated by an excitation located solely on two qubits in the middle of the spin chain at $t=0$ the following matchgate circuit can be used:
\begin{enumerate}[(i)]
	\item	Prepare the initial state $\ket{\Psiin}=\ket{0}^{\otimes n}$, i.e the ground state of  $H(0)=-B H_0$;
	\item	evolve the system adiabatically, by applying the unitary $U_T^{(1)}(J)$ in Eq. (\ref{eq:UT1}), for a certain value of $J$ to obtain the ground state of $H(J)$ from Eq. (\ref{eq:Hamiltonian simplified});
	\item	create the signal by applying the unitary $U_T^{(2)}$ in Eq. (\ref{eq:UT2}) in order to flip the spin component of a pair of spins;
	\item evolve the state by the time evolution unitary $U_T^{(3)}(J,t)$ in Eq. (\ref{eq:UT3fin}) for a certain time $t$.
	\item	measure the $k$-th qubit in the $z$ direction to obtain the expectation value $\bk{Z_k(J,t)}$;
	\item	repeat the previous steps for all $k$ to obtain the $z$ spin component of every spin in the chain;
	\item	repeat the whole procedure for different values of $t$ between $0$ and some $t_{\mathrm{max}}$ to measure the propagation in time of the signal along the chain.
\end{enumerate}

\subsection{Compressed circuit for time evolution}\label{sec:timeevolution-compressed}

In the following, we construct a compressed circuit to compute 	$\bk{Z_k(J,t)}$ in Eq. (\ref{eq:Zk(J,t)}). Below we compute the matrix $R_T(J,t)$ associated to $U_T(J,t)$. As before, we use that $Z_k=-ic_{2k-1}c_{2k}$ and Eq. (\ref{eq:UcU}) to write Eq. (\ref{eq:Zk(J,t)}) as
\be
	\ba
		\bk{Z_k(J,t)}
		&={\bmk{\Psiin}{U^{\dagger}_T(J,t)(-ic_{2k-1}c_{2k})U_T(J,t)}{\Psiin}}	\\
		&=\bmk{2k-1}{R_T(J,t)SR_T^T(J,t)}{2k}\\
		&=\tr{\left[R_T^T(J,t)\rhoin^{(k)}R_T(J,t)Y_m \right]},
	\ea
	\label{eq:Zk}
\ee
where we have used that $S=iY_m$. The initial state, $\rhoin^{(k)}$ is given by $ \rhoin^{(k)}=\frac{1}{2}\big(\ket{2k-1}+i\ket{2k}\big)\big(\bra{2k-1}-i\bra{2k}\big)$, which can be written as
\be
	\rhoin^{(k)}=\kb{k}{k}\otimes \kb{+_{y}}{+_{y}}_m.
	\label{eq:rho_ink}
\ee

Equation (\ref{eq:Zk}) shows that the expectation value of the operator $Z_k$ (as a function of time) can be measured by the following circuit:
\begin{enumerate}[(i)]
	\item	Prepare the $m$--qubit initial state $\rhoin^{(k)}$ in Eq. (\ref{eq:rho_ink});
	\item	evolve the system by applying the real operator $R_T(J,t)$ [computed below, see Eq. (\ref{eq:RT})] for certain values of $J$ and $t$;
	\item	measure the last qubit in the $y$--basis to obtain the expectation value  $\bk{Z_k(J,t)}$ in Eq. (\ref{eq:Zk});
	\item	repeat the process for every $k$ to obtain the $z$ spin component of every spin in the chain at time $t$;
	\item	repeat the whole procedure for different values of $t$ between $0$ and some $t_{\mathrm{max}}$ to measure the propagation of the signal along the chain as a function of time.
\end{enumerate}

In the Fig. \ref{fig:signal propagation} we show the propagation observed with the classical simulation of the compressed algorithm, and in Fig.  \ref{fig:propagation speed} we illustrate the relation between the signal propagation speed and the parameters $J$ and $\delta$.

In contrast to the circuits studied before, it is not possible here to further compress this circuit [to $\log(n)$ qubits] using the same tools as before. The reason for that is that the state $\rhoin^{(k)}$ does not transform into a direct sum [see Eq. (\ref{eq:block diagonals})] under conjugation by the operator $V$ in Eq. (\ref{eq:V}).

\begin{figure}[h]
  \centering
  \includegraphics[width=0.23\textwidth]{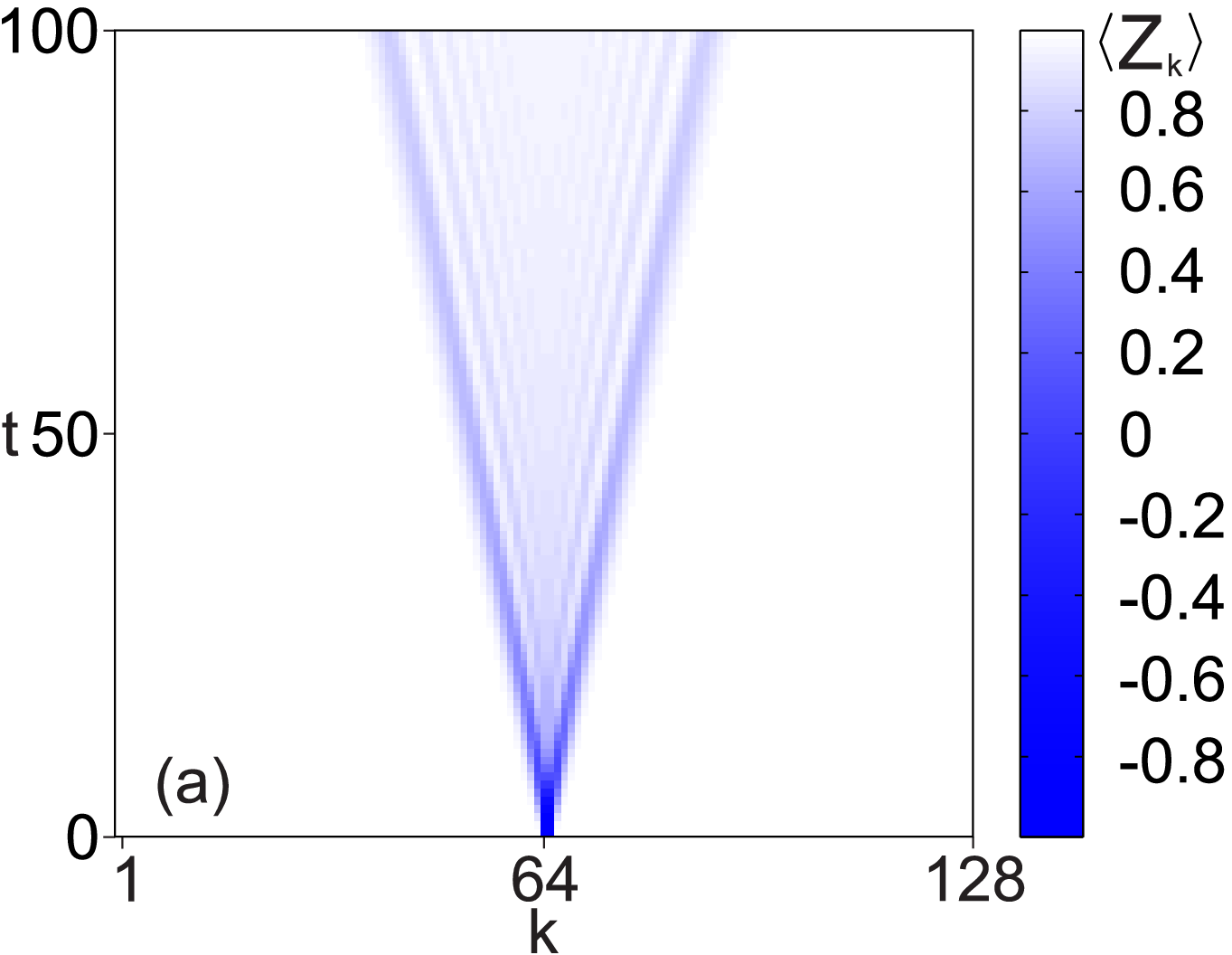}
  \includegraphics[width=0.23\textwidth]{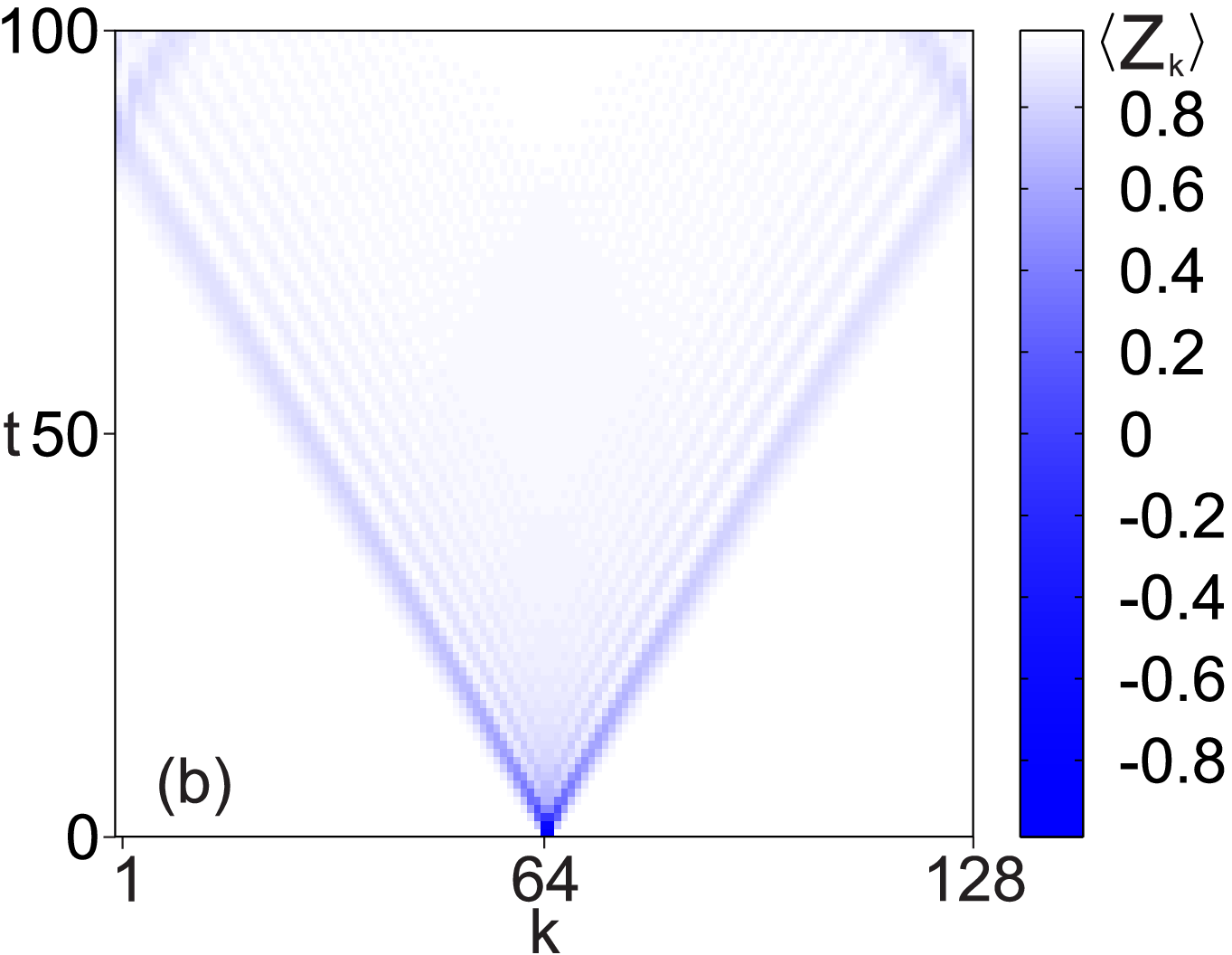}
  \includegraphics[width=0.23\textwidth]{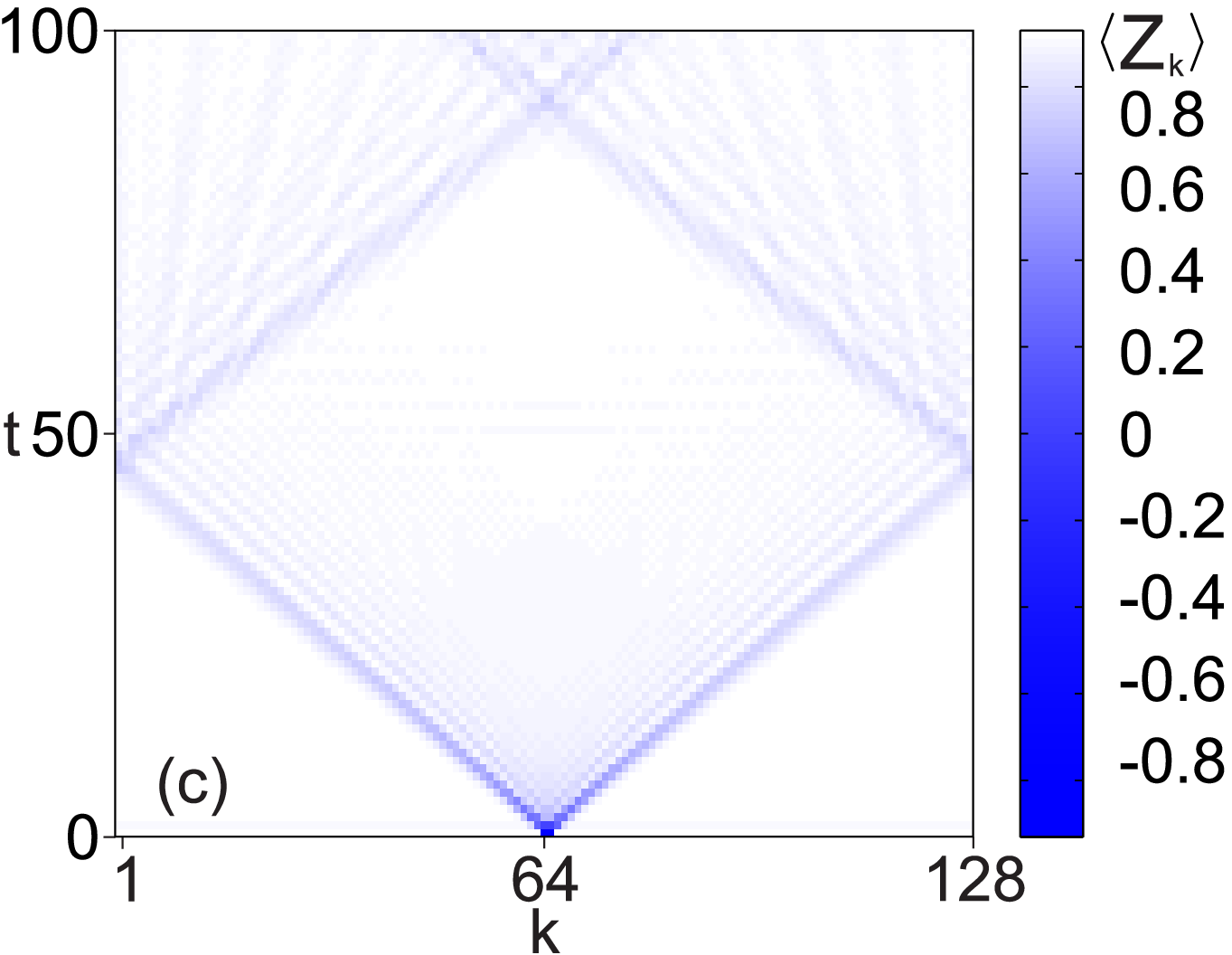}
  \caption{Propagation of an excitation generated in the middle of a 1D chain of $n=128$ spins, with interactions given by a XY Hamiltonian with JW boundary conditions, depicted for $J_{max}=0.1, 0.3$ and $0.6$ in (a), (b) and (c) respectively. The expectation values $\bk{Z_k(J,t)}$ for every spin $k$ were obtained via the circuit of Sec. \ref{sec:timeevolution-compressed} [see Eq. (\ref{eq:Zk})].}
  \label{fig:signal propagation}
\end{figure}

\begin{figure}[h]
  \centering
  \includegraphics[width=0.4\textwidth]{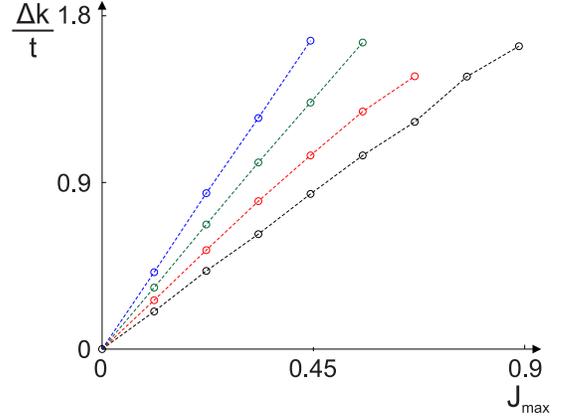}
  \caption{Propagation speed, $\frac{\Delta k}{T}$, as a function of $\Jmax$ obtained with the circuit presented in Sec. \ref{sec:timeevolution-compressed}. The different curves corresponds from bottom to top to values of $\delta=0,0.3,0.6$ and $0.9$.}
  \label{fig:propagation speed}
\end{figure}

\subsubsection*{Derivation of the matrix $R_T$}

Here we construct the matrices $R_T(J,t)$ associated to $U_T(J,t)$. Since $U_T^{(1)}=U(J)$ in equation (\ref{eq:U=U0U1U2}),
\be
	R_T^{(1)}(J)=R(J),
\ee
where $R(J)$ is given by equation (\ref{eq:R=R0R1R2}). To construct the matrix $R_T^{(2)}$ associated to the unitary $U_T^{(2)}$, we write
\be
	U_T^{(2)}=e^{i\pi H_T^{(2)}/2},
\ee
where the Hamiltonian $H_T^{(2)}=-ic_n c_{n+1}$. According to Eq. (\ref{eq:Quadratic Hamiltonian}), the matrix $h^{(2)}$ associated to this Hamiltonian has only two non-zero elements $h^{(2)}_{n,n+1}=-1/2$ and $h^{(2)}_{n+1,n}=1/2$. This implies that the matrix $R^{(2)}=e^{-2\pi h^{(2)}}$ associated to $U_T^{(2)}$ is a diagonal matrix with all diagonal elements equal to one, except the elements
\be
	  [R_T^{(2)}]_{n,n}=[R_T^{(2)}]_{n+1,n+1}=-1.
\ee

The matrix $R_T^{(3)}(J,t)$, associated to $U_T^{(3)}(J,t)$, is obtained using Eq. (\ref{eq:UT3fin}) as
\be
	R_T^{(3)}(J,t)=\big\{R_0[\zeta_0(t)]R_1[\zeta_1(J,t)]R_2[\zeta_2(J,t)]\big\}^{L_T},
\ee
where $R_0$, $R_1$ and $R_2$ are given by Eq. (\ref{eq:R0R1R2}). The matrix which has to be used in Eq. (\ref{eq:Zk}) is therefore
\be
	R_T(J,t)=R_T^{(3)}(J,t)R_T^{(2)}R_T^{(1)}(J).
	\label{eq:RT}
\ee

\section{Conclusion and outlook}

In summary, we have investigated here several physically relevant processes, which can be realized with a universal quantum computer operating on very few qubits. The reason why this compressed way of quantum simulation works is because all the circuits investigated here where matchgate circuits, for which it has been shown that their power coincides with a universal quantum computer of exponentially smaller width. It should be noted that any computation which can be simulated in the strong sense by an exponentially smaller system, as it is done here, must be classically efficiently simulatable since the dimension of the Hilbert space describing the system is linear in $n$. Regarding classical simulation one distinguishes between a strong simulation, and a weak simulation \cite{vN07}. Strong simulation means that the probabilities of the measurement outcomes is computed efficiently exactly, whereas weak simulation means that one can sample from this probability distribution classically efficiently. Those two notions are fundamentally different, and quantum computations which cannot be simulated strongly might well be weakly simulatable \cite{vN07}. Note that in the compressed simulation considered here, the probabilities of the measurement outcomes of both, the circuits of width $n$ and the one of width $\log(n)$ coincide.

We intent to generalize the notion of compressed quantum computation to other models, like for instance the 6--vertex model. Even though it might not be possible to simulate such a model quantum mechanically in the strong sense, as it was done for matchgate circuits, their weak simulation might be feasible.

Another interesting problem, which we intend to address in future is based on the second part of Theorem 2 (see Sec. \ref{sub: matchgates and theorems}). In particular, it would be interesting to investigate the scenario where a perturbation is added to the compressed quantum computer. Even though the realization of this perturbation might be rather straight forward on the small system, it might be difficult to analyze it with matchgates. This would imply that its classical simulation based on the Jordan Wigner transformation might no longer be feasible. We anticipate that in this case the classical simulation might in general be hard.

\appendix

\section{exact diagonalization of the Ising and XY model}\label{ap:diagonalization}

The exact diagonalization of the Hamiltonian
\begin{equation} \label{eqn:XY}
H = -B \sum_{i=1}^n -J \sum_{i=1}^{n} (X_i X_{i+1} + \delta Y_i Y_{i+1})
\end{equation}
is performed via Jordan-Wigner transformation \cite{jordanwigner28,lieb61}
which consists in a mapping of the set of Pauli-operators on a set of
fermion creation and annihilation operators.

In order to specify the Jordan-Wigner transformation it is advantageous to
introduce the raising and lowering operators
\begin{displaymath}
    \adj{a_i} = \frac{1}{2} ( X_i + i Y_i )
    \textrm{ and }
    a_i = \frac{1}{2} ( X_i - i Y_i ).
\end{displaymath}
These operators partly obey the commutation relations of Bose
operators and they partly obey the anticommutation relations of
Fermi operators:
\begin{eqnarray*}
    [a_i, \adj{a_j}]=0, & [\adj{a_i}, \adj{a_j}] = [a_i, a_j] = 0 &
        \textrm{ for } i \neq j \\
    \{a_i, \adj{a_i} \} = 1, & \{\adj{a_i}, \adj{a_i}\} = \{a_i, a_i\} =
        0 & \\
\end{eqnarray*}
In terms of the raising and lowering operators the Pauli operators
can be written as
\begin{eqnarray*}
    X_i & = & \adj{a_i} + a_i \\
    Y_i & = & \frac{1}{i} ( \adj{a_i} - a_i ) \\
    Z_i & = & 2 \adj{a_i} a_i - 1
\end{eqnarray*}
and the Hamiltonian (\ref{eqn:XY}) equals
\begin{eqnarray*} \label{eqn:jw:hisinga}
    H  & = & - J \sum_{i=1}^n [ (1-\delta) \big( \adj{a_i} \adj{a_{i+1}} + a_i a_{i+1} \big) + \nonumber\\
       & & \phantom{J \sum_{i=1}^n} + (1+\delta) \big( a_i \adj{a_{i+1}} + \adj{a_i} a_{i+1} \big) ] -\\
       & & - 2 B \sum_{i=1}^n \adj{a_i} a_i + B n\\
\end{eqnarray*}
The Hamiltonian is a quadratic form in the raising and lowering
operators $\adj{a_i}$ and $a_i$. Unfortunately, it is not possible
to find a linear transformation of the operators $\adj{a_i}$ and
$a_i$ that diagonalizes this Hamiltonian and preserves the
commutation and anticommutation relations.
The Jordan-Wigner transformation
\begin{eqnarray} \label{eqn:jw:jwtrafo}
    c_i  =  e^{i \pi \sum_{j=1}^{i-1} \adj{a_j} a_j} \, a_i , & & i \in
    \{1,\ldots,n\}
\end{eqnarray}
transforms the set of raising and lowering operators to a set of
real Fermi operators, in terms of which the Hamiltonian is still a
quadratic form (up to a boundary term that depends on the parity),
namely
$H = H_{quad} + H_{bc}$
with
\begin{eqnarray} \label{eqn:XYc}
    H_{quad}  & = & - J \sum_{i=1}^n [ (1-\delta) \big( \adj{c_i} \adj{c_{i+1}} + c_i c_{i+1} \big) + \nonumber\\
       & & \phantom{J \sum_{i=1}^n} + (1+\delta) \big( c_i \adj{c_{i+1}} + \adj{c_i} c_{i+1} \big) ] -\nonumber\\
       & & - 2 B \sum_{i=1}^n \adj{c_i} c_i + B n
\end{eqnarray}
and
\begin{eqnarray*}
    H_{bc}  & = & - J [ (1-\delta) \big( \adj{c_n} \adj{c_1} + c_n c_1 \big) + \nonumber\\
       & & \phantom{J} + (1+\delta) \big( c_n \adj{c_1} + \adj{c_n} c_1 \big) ] (1 + e^{i \pi \hat{N}}).
\end{eqnarray*}
Here, $\hat{N}=\sum_{i=1}^n \adj{c_i} c_i$ denotes the fermion number-operator
and $c_{n+1} = c_1$.

The boundary term $H_{bc}$ guarantees the compliance with the periodic boundary conditions.
It is not quadratic due to the parity operator $e^{i \pi \hat{N}}$.
However, the parity is a symmetry of the Hamiltonian. As such, it
simplifies to a c-number in the subspace of even- and odd number of fermions.
Because of this, even- and odd number of particles have to be treated
separately in case of periodic boundary conditions.

The effect of the correction term $H_{bc}$ on the
Eigenvalues and Eigenstates of the Hamiltonian is of order $1/N$.
Thus, for the calculation of real physical quantities of large
systems ($N \gg 1$) the effect of the correction term is
negligible. It is therefore justified to ignore the correction
term $H_{bc}$ and deal with the simplified Hamiltonian $H \approx H_{quad}$.
The neglect of the correction term $H_{bc}$ equals the assumption
of very special boundary conditions, given by
\begin{eqnarray}
    \label{eqn:jw:jwbc}
    X_{n+1} & = &
    \left( \prod_{i=1}^n Z_i \right) X_1,\\
    Y_{n+1} & = &
    \left( \prod_{i=1}^n Z_i \right) Y_1 \textrm{ and }\nonumber\\
    Z_{n+1} & = & i Y_{n+1} X_{n+1}.\nonumber
\end{eqnarray}
These boundary conditions will be referred to as ``Jordan-Wigner
boundary conditions''. The Hamiltonian (\ref{eqn:XY})
therefore no longer describes a cyclic chain that is invariant
under translations, but an open chain the last spin of which
couples to the operator (\ref{eqn:jw:jwbc}).
With Jordan-Wigner boundary conditions, the system is periodic
in terms of the operators $c_i$, i.e. it is invariant under
the translation $c_i \to c_{i+1}$.
This invariance gives rise to another symmetry of the Hamiltonian,
namely the conservation of the momentum
\begin{displaymath}
\hat{P} = -i \sum_i \left( \adj{c_j} c_{j-1} - \adj{c_j} c_{j+1} \right).
\end{displaymath}

In case of open boundary conditions, the Hamiltonian $H$ equals $H_{quad}$
with the first sum going from $1$ to $n-1$ instead of $n$.
Since there is no boundary term, the Hamiltonian is a quadratic
form in terms of Fermi operators.
It conserves parity, but not momentum.

The symmetries of the Hamiltonian, i.e. the parity conservation and,
in case of Jordan-Wigner boundary conditions, the momentum conservation,
give rise to level-crossings in the spectrum.
These level-crossings take place between states of different symmetry,
as can be gathered from Figs.~\ref{fig:spectrum open bc}
and~\ref{fig:spectrum JW bc}.
\begin{figure}[h!]
  \centering
  \includegraphics[width=0.23\textwidth]{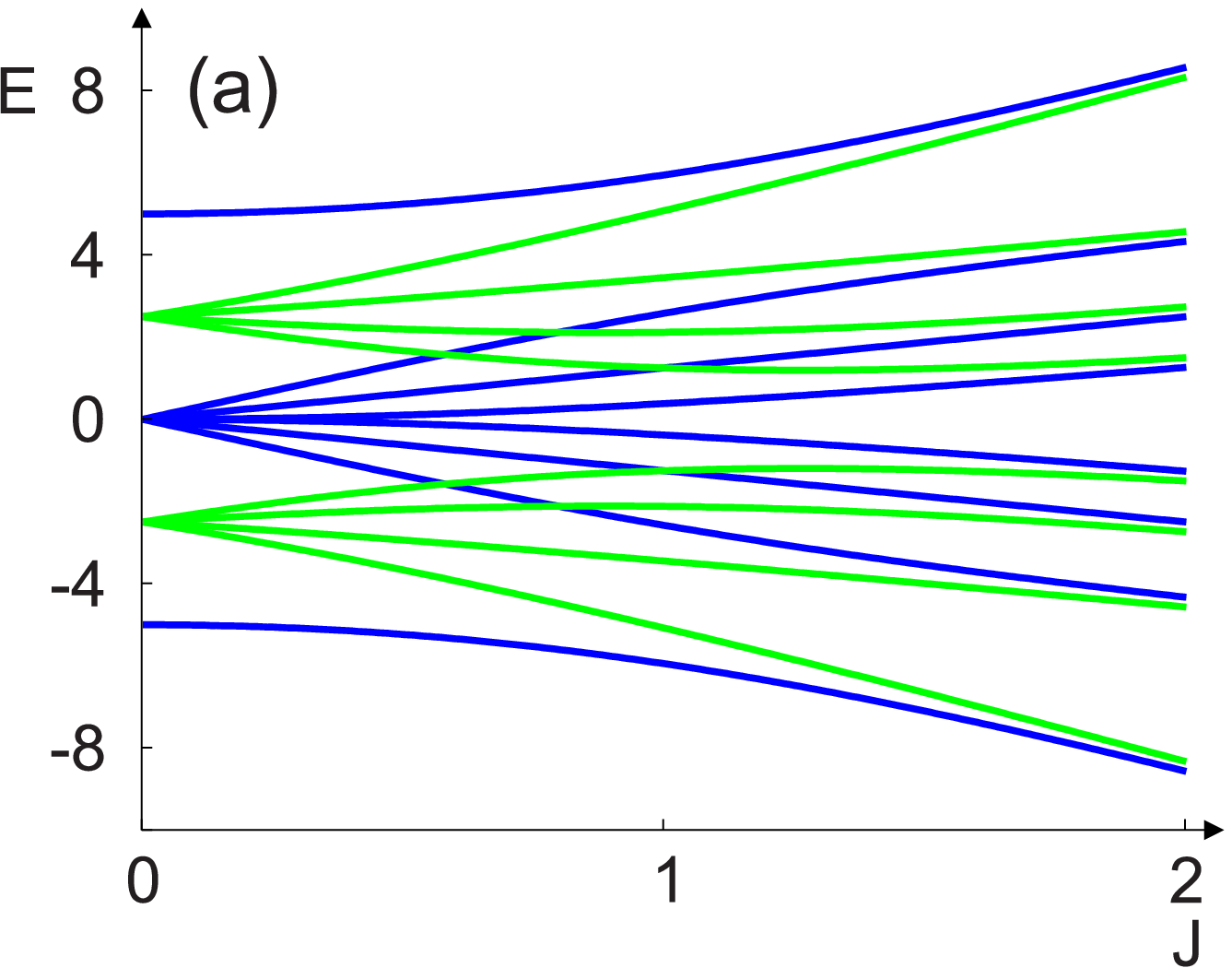}
  \includegraphics[width=0.23\textwidth]{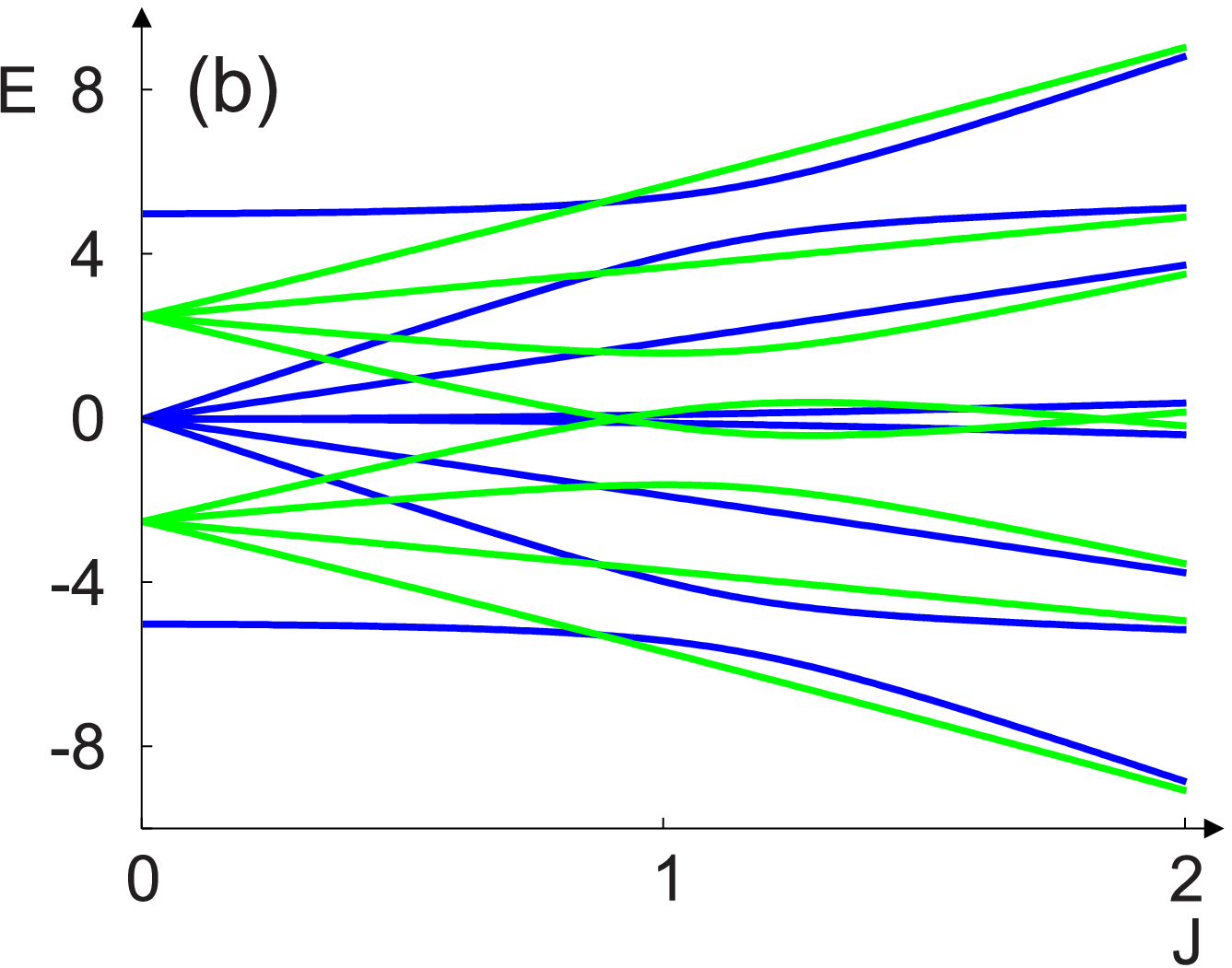}
  \includegraphics[width=0.23\textwidth]{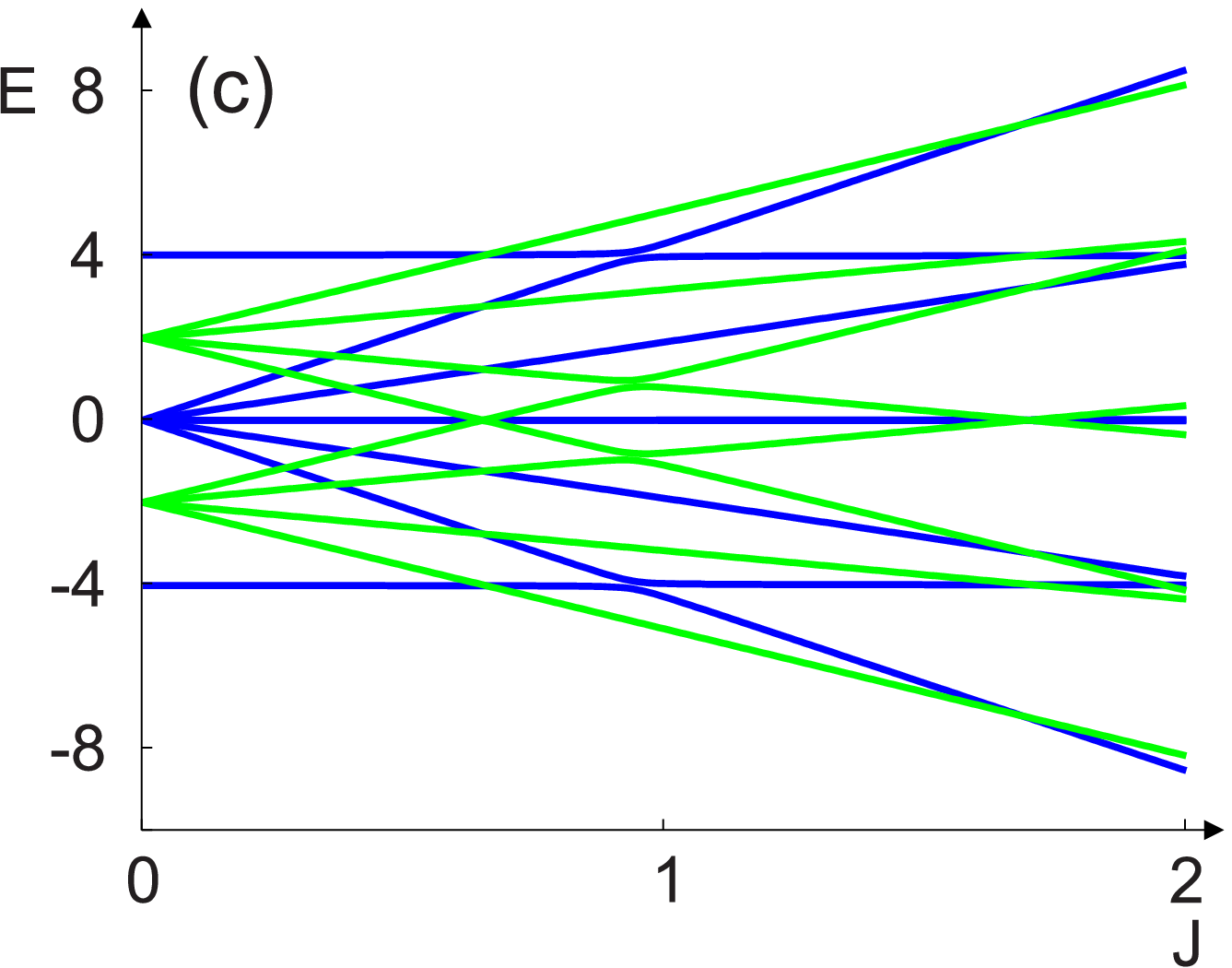}
  \caption{Energy spectrum of the XY Hamiltonian with open boundary conditions, computed analytically for $n=4$. The parameter $\delta$ was set to $0$, $0.5$ and $0.9$ in (a),(b) and (c) respectively. Different colors indicate energies corresponding to eigenstates of different parity. Note that a system prepared in the ground state for $J=0$ will not experience any level crossing during an adiabatic evolution, since the parity is preserved.}
  \label{fig:spectrum open bc}
\end{figure}

\begin{figure}[h]
  \centering
  \includegraphics[width=0.23\textwidth]{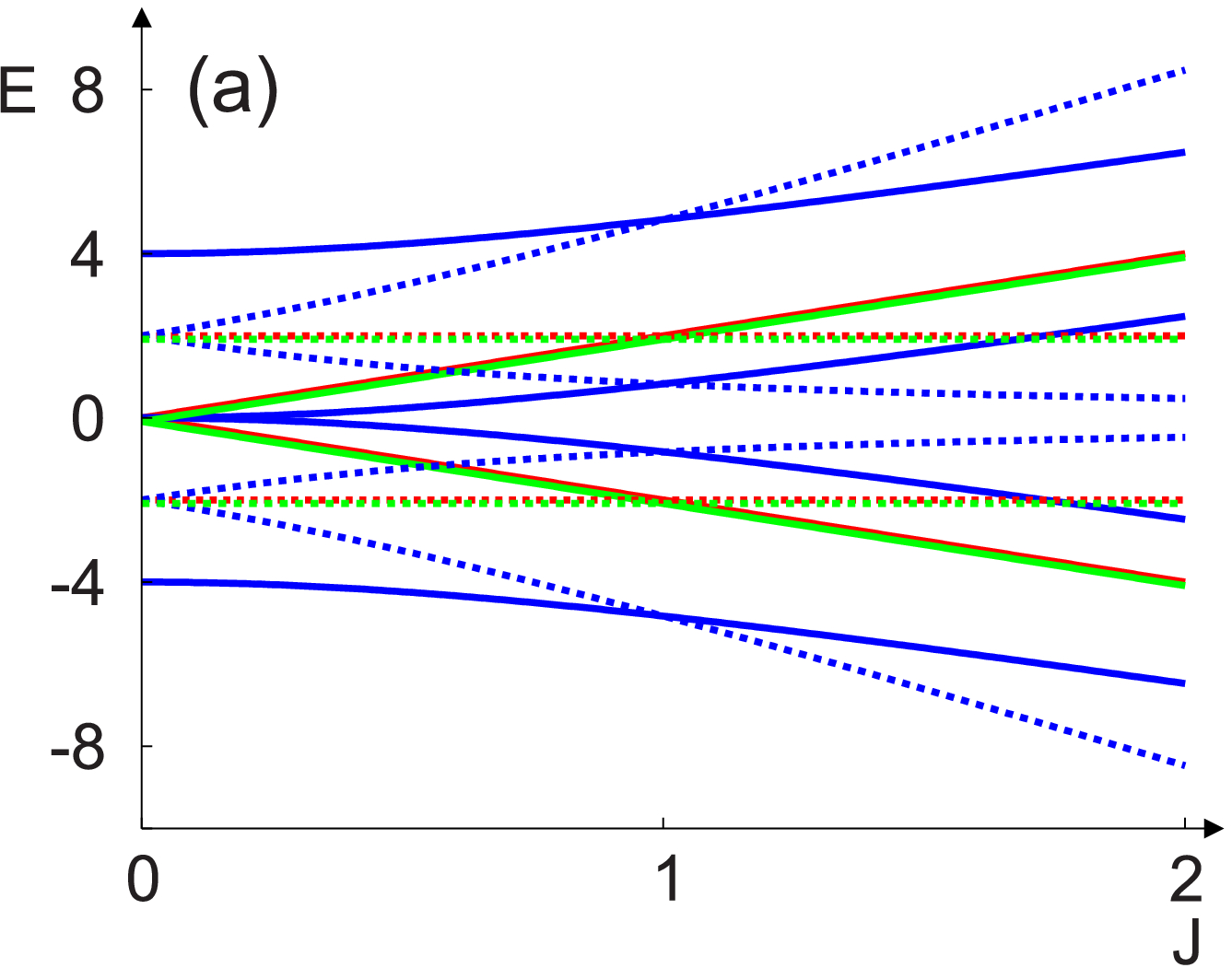}
  \includegraphics[width=0.23\textwidth]{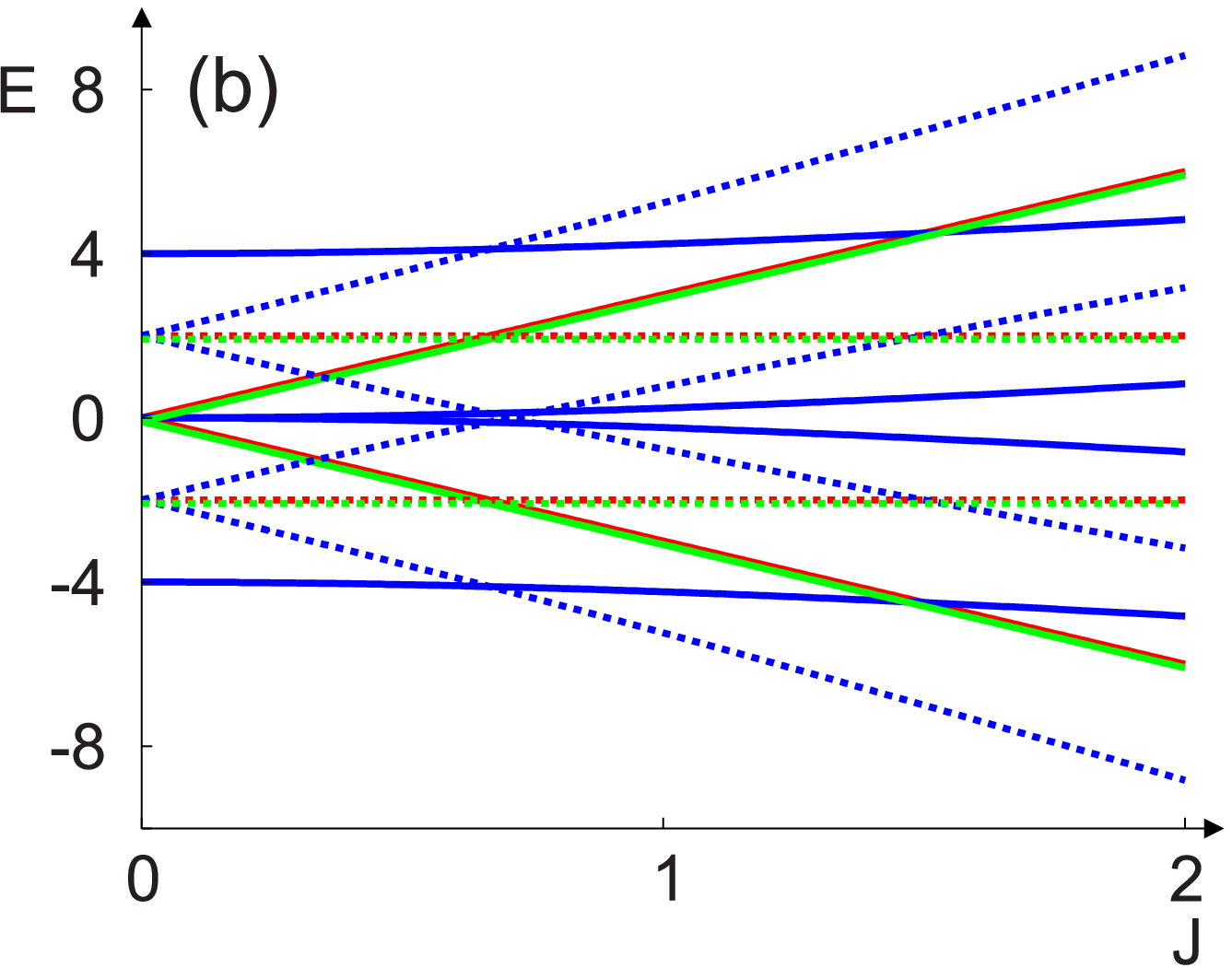}
  \includegraphics[width=0.23\textwidth]{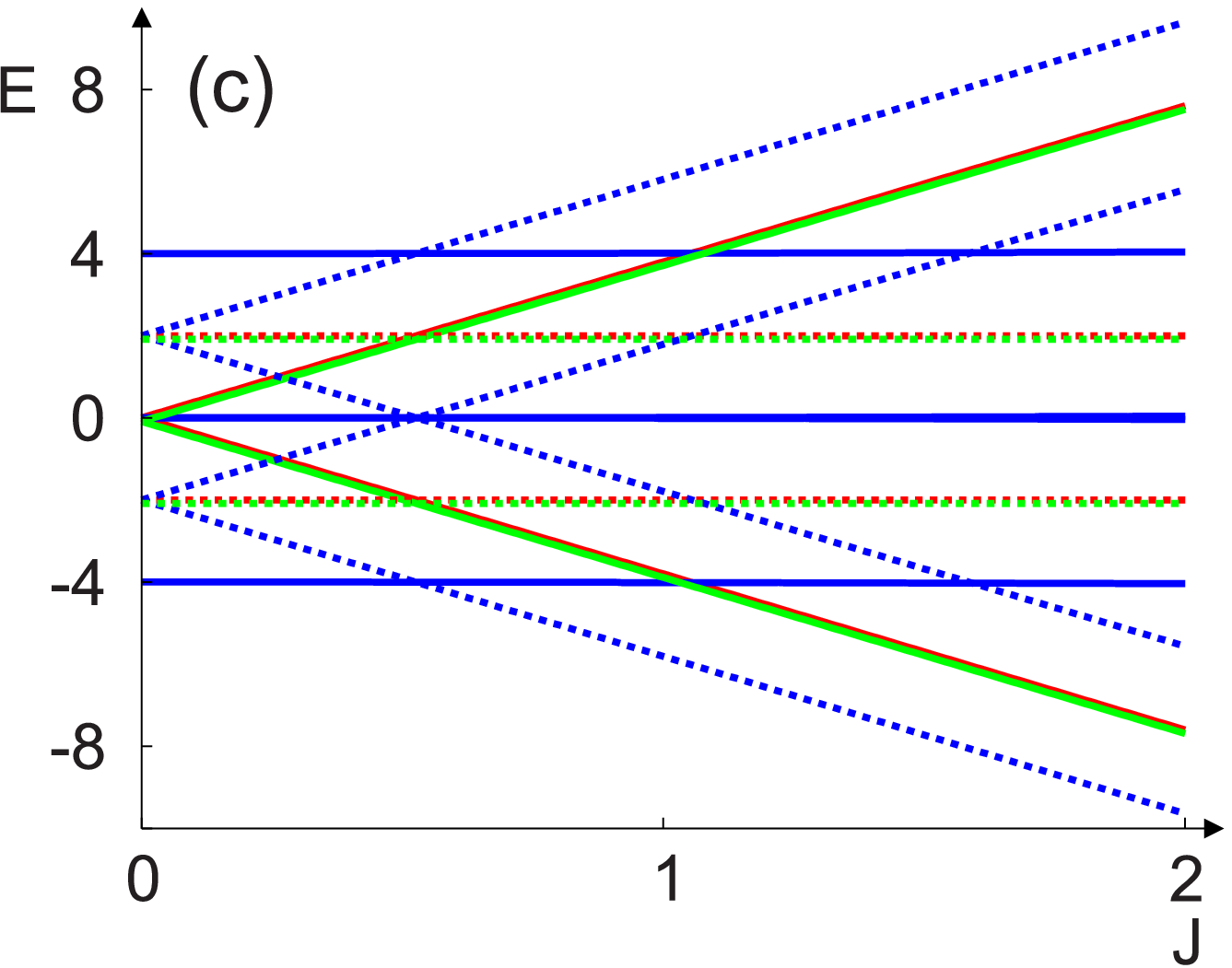}
  \caption{Energy spectrum of the XY Hamiltonian with JW boundary conditions, computed analytically for $n=4$. The parameter $\delta$ was set to $0$, $0.5$ and $0.9$ in (a),(b) and (c) respectively. We indicate with solid lines the energies corresponding to eigenstates of parity $+1$ and with dashed lines the energies corresponding to eigenstates of parity $-1$. Different colors distinguish energies corresponding to eigenstates with different momentum. Note that a system prepared in the ground state for $J=0$ will not experience any level crossing during an adiabatic evolution, since the parity and momentum are preserved.}
  \label{fig:spectrum JW bc}
\end{figure}
Irrespective of the boundary conditions, the Hamiltonian
will be a quadratic form in terms of the Fermi operators~$c_i$.
Diagonalization of a quadratic form can be performed
in $poly(N)$ steps as described in~\cite{lieb61}.
The idea is to find a Bogoliubov transformation
\begin{equation} \label{eqn:bog}
\eta_k = \sum_i \left( g_{k i} c_i + h_{k i} \adj{c_i} \right)
\end{equation}
such that the Hamiltonian is diagonal in term of the new Fermi operators $\eta_k$:
\begin{equation} \label{eqn:xyeta}
H = E_0 + \sum_k \Lambda_k \adj{\eta_k} \eta_k.
\end{equation}

The general form of a quadratic Hamiltonian is
\begin{displaymath}
H = \sum_{i j} \left[ \adj{c_i} A_{i j} c_j + \frac{1}{2} \left( \adj{c_i} B_{i j} \adj{c_j} + h.c. \right) \right] + const
\end{displaymath}
with $A$ being real and symmetric and $B$ being real and antisymmetric.
For the XY model with open boundary conditions,
$A_{i i}=-2 B$, $A_{i,i+1}=A_{i+1,i}=-J(1+\delta)$,
$B_{i,i+1}=-B_{i+1,i}=-J(1-\delta)$, $i=1,...,n-1$ and $const = B n$.
The XY model with Jordan Wigner boundary conditions has in addition
the elements $A_{n,1}$ and $A_{1,n}$ set to $-J(1+\delta)$
and $B_{n,1}$ and $B_{1,n}$ set to $-J(1-\delta)$ and
$J(1-\delta)$ respectively.

The constant part~$E_0$ in~(\ref{eqn:xyeta}) can be determined from
the property that the trace of~$H$ is invariant under the
Bogoliubov transformation~(\ref{eqn:bog}). In this way, $E_0$ is obtained as

\begin{displaymath}
E_0 = \frac{1}{2} \left( \sum_i A_{i i} - \sum_k \Lambda_k \right).
\end{displaymath}

The energies $\Lambda_k$ of the $\eta$-fermions and the coefficients
$g_{k i}$ and $h_{k i}$ of the linear transformation are obtained
by solving two $N \times N$ eigenvalue problems.
Plugging the transformation~(\ref{eqn:bog}) in
the Hamiltonian, it can be gathered that the diagonal form~(\ref{eqn:xyeta})
is achieved if the equations
\begin{eqnarray} \label{eqn:gheq}
g A - h B & = & \Lambda g \\
g B - h A & = & \Lambda h \nonumber
\end{eqnarray}
are fulfilled
with $\Lambda$ being the diagonal matrix $\Lambda_{k l} = \Lambda_k \delta_{k l}$.
These equations guarantee that the terms $\adj{\eta_k} \eta_l$
disappear for $k \neq l$.
The terms $\eta_k \eta_l$ and $\adj{\eta_k} \adj{\eta_l}$ vanish
provided that
\begin{equation} \label{eqn:gheq2}
[h \adj{g},\Lambda]=0.
\end{equation}
Equations~(\ref{eqn:gheq}) imply the two
eigenvalue problems
\begin{eqnarray*}
G_{k-} (A-B)(A+B) & = & \Lambda_k^2 G_{k-} \\
H_{k-} (A+B)(A-B) & = & \Lambda_k^2 H_{k-}.
\end{eqnarray*}
with $G=g+h$ and $H=g-h$.
The eigenvalues $\Lambda_k^2$ are positive, such that
$\Lambda_k$ is real and fixed up to the sign.
The signs of the $\Lambda_k$'s can be gathered from condition~(\ref{eqn:gheq2}).

The ground state is the Fermi-see of
$\eta$-fermions with negative energy:
$\ket{\Psi_0} = \adj{\eta_1} \cdots \adj{\eta_{k_F}} \ket{0_{\eta}}$
if $\Lambda_1,\ldots,\Lambda_{k_F} < 0$.
The parity of the ground state equals the parity of the
number of $\eta$-fermions required to build the ground state.
The first excited states
are the single-particle or single-hole excitations around the
Fermi-level~$k_F$, i.e.
$\adj{\eta_{k_F+1}} \ket{\Psi_0}$ or
$\eta_{k_F} \ket{\Psi_0}$.
These states have different parity than the ground state.
The first states with the same parity as the ground state
are the two-particle excitation
$\adj{\eta_{k_F+1}} \adj{\eta_{k_F+2}} \ket{\Psi_0}$, the two-hole excitation
$\eta_{k_F-1} \eta_{k_F} \ket{\Psi_0}$ or the particle-hole excitation
$\adj{\eta_{k_F+1}} \eta_{k_F} \ket{\Psi_0}$.
An avoided level-crossing
between these states and the ground state is an indication for a
quantum phase transition (see Fig. \ref{fig:energy gap}).
\begin{figure}[h!]
  \centering
  \includegraphics[width=0.4\textwidth]{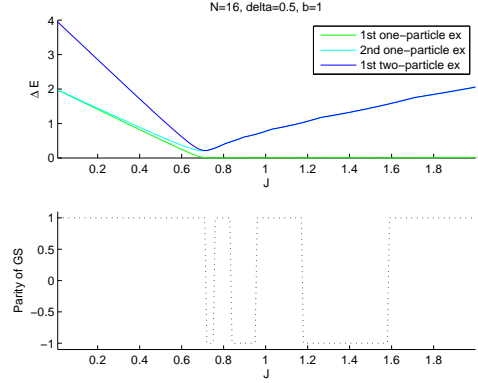}
  \caption{In the upper plot the energy gap with respect to the ground state energy of the single-particle, single-hole and two-particle excitation states are depicted. In the lower plot, the parity of the groundstate is indicated. Note that every level-crossings in the ground state coincides with a change of the parity in the ground state.}
  \label{fig:energy gap}
\end{figure}

\begin{figure}[h]
  \centering
  \includegraphics[width=0.23\textwidth]{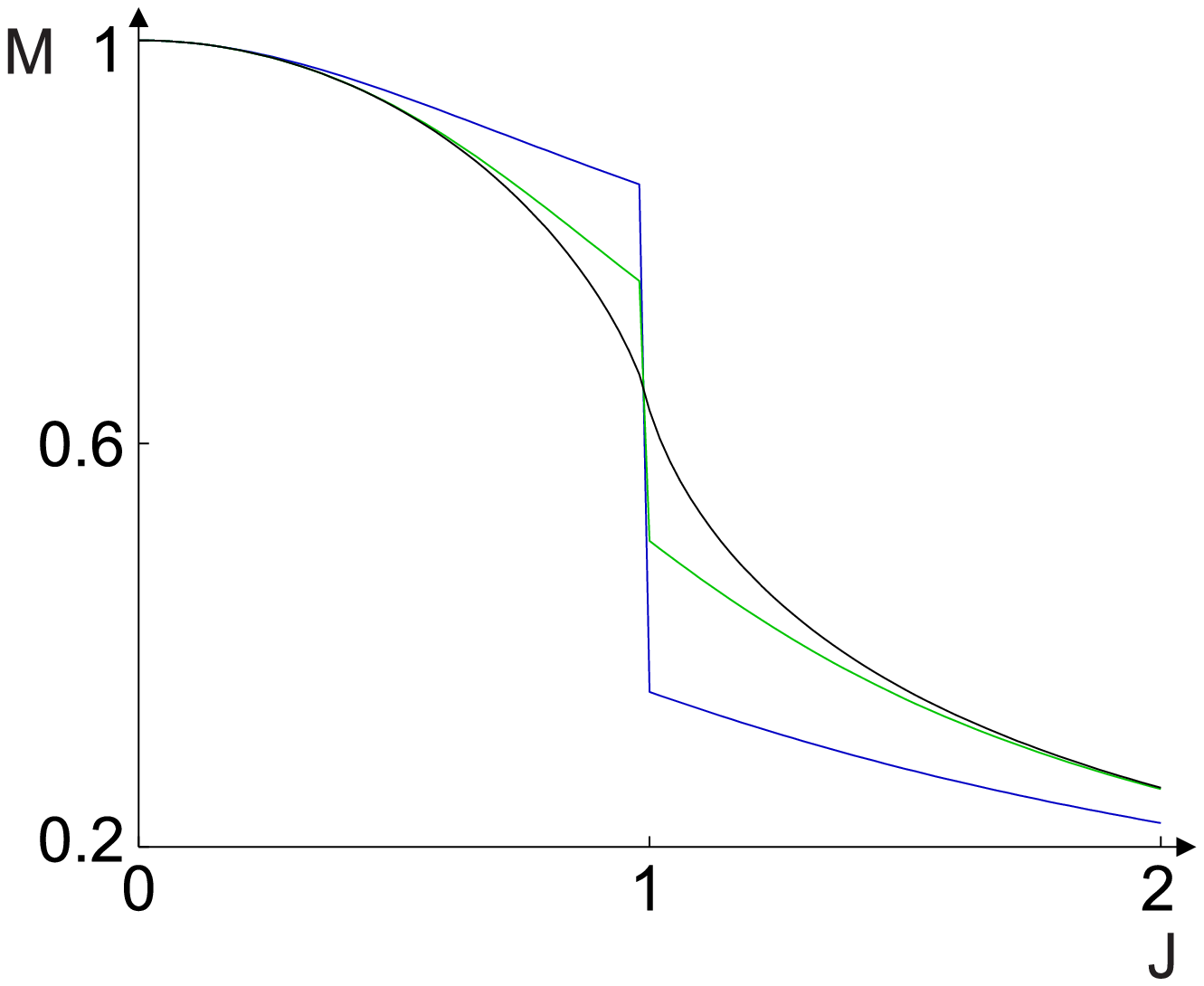}
  \includegraphics[width=0.23\textwidth]{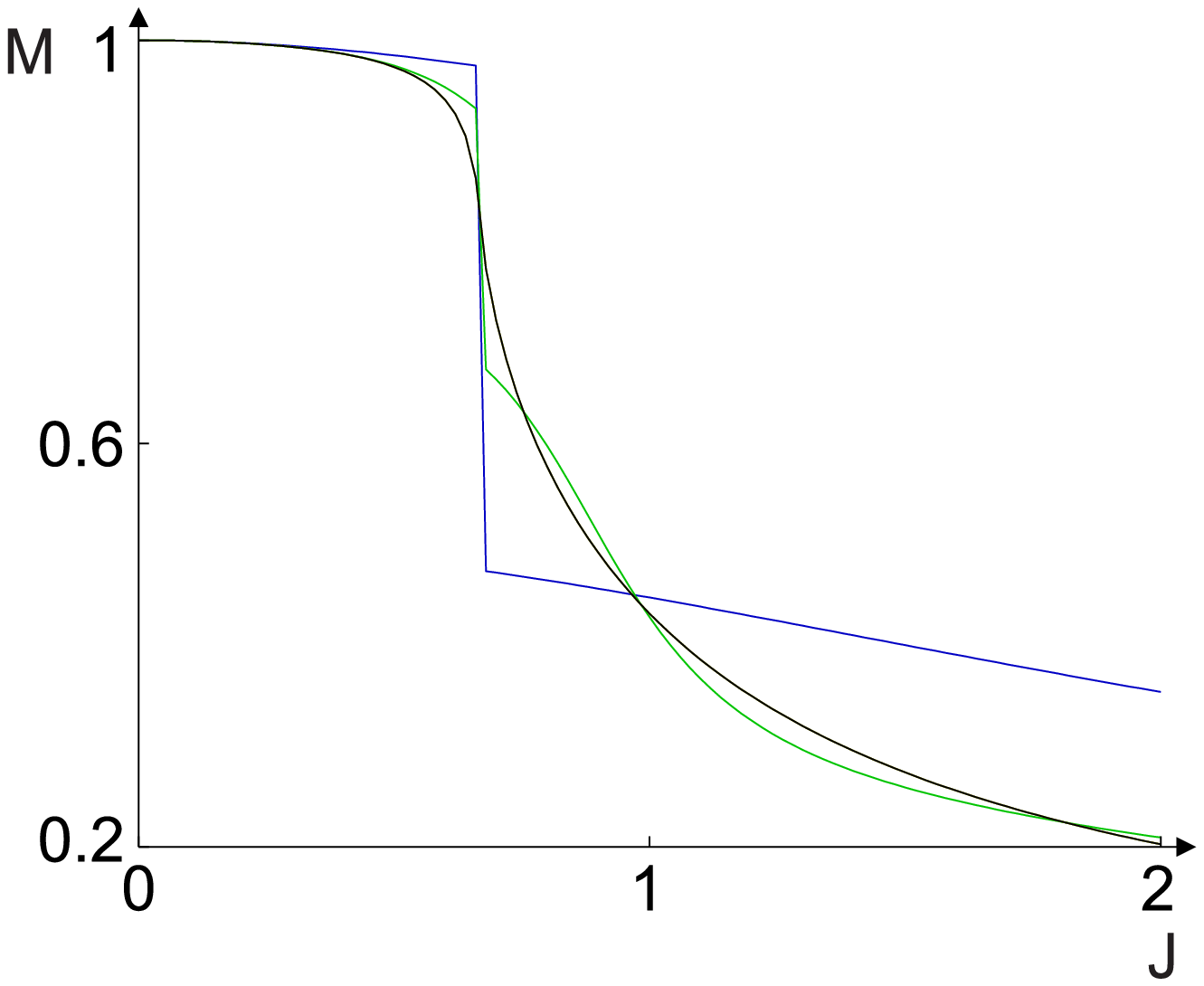}
  \caption{Magnetization of the ground state of the XY Hamiltonian with JW boundary conditions, for $\delta=0$ (left plot) and $\delta=0.5$ (right plot). Different curves correspond to system sizes of $n=4,8$ and $256$ respectively. The gap decreases with $n$ to the point that, for $n$ sufficiently large, this curve coincides with the magnetization curve that can be obtained with the circuits in Sec. \ref{sec:magnetization} (see Fig. \ref{fig:magnetization-adiabaticevolution}).}
  \label{fig:magnetization-groundstate}
\end{figure}

The magnetization in $z$-direction, $M_z = \frac{1}{n} \sum_{i=1}^n \expect{Z_i}$,
of the ground state is obtained as
\begin{displaymath}
M_z = \frac{2}{n} \sum_{i=1}^n
\left(
\sum_{k=1}^{n} h_{k i}^2 +
\sum_{k=1}^{k_F} (g_{k i}^2 - h_{k i}^2)
\right) - 1.
\end{displaymath}
This is due to the identity $Z_i = 2 \adj{c_i} c_i -1$
and the fact that
$\expect{\adj{c_i} c_i} = \sum_k \left( h_{k i}^2 + (g_{k i}^2 - h_{k i}^2) \expect{\adj{\eta_k} \eta_k}\right)$. In Fig. \ref{fig:magnetization-groundstate}, we plot this function for different system sizes and boundary conditions. As explained above, the ground  state does not have a unique parity or momentum for every value of $J$, which produces a discontinuity in the magnetization function, as can be seen in Fig. \ref{fig:magnetization-groundstate}.

\end{document}